\documentclass[10pt,journal]{IEEEtran}%
\ifCLASSOPTIONcompsoc
  \usepackage[nocompress]{cite}
\else
  \usepackage{cite}
\fi
\usepackage{varwidth}
\ifCLASSINFOpdf
\usepackage[pdftex]{graphicx}
\else
\fi

\usepackage{amsfonts}
\usepackage{amsmath}
\usepackage{amssymb}
\usepackage{bbm}
\usepackage{xcolor}
\newcommand{\RNum}[1]{\uppercase\expandafter{\romannumeral #1\relax}}
\usepackage{hyperref}

\usepackage{graphicx}
\usepackage{float}
\usepackage[caption=false]{subfig}
\usepackage{upgreek}
\usepackage[ruled,vlined]{algorithm2e}

\usepackage{amsthm}

\newcommand{\rev}[1]{{#1}} %
\newcommand{\ignore}[1]{}

\begin{document}
\title{Broadband Digital Over-the-Air Computation for Wireless Federated Edge Learning}

\author{Lizhao~You,~%
		Xinbo~Zhao,~%
        Rui~Cao,
        Yulin~Shao,~%
        and~Liqun~Fu%
\IEEEcompsocitemizethanks{\IEEEcompsocthanksitem L. You, X. Zhao, R. Cao, and L. Fu are with the Department
of Information and Communication Engineering, School of Informatics, Xiamen University, China. %
E-mails: \{lizhaoyou, liqun\}@xmu.edu.cn, \{xbzhao, ruicao\}@stu.xmu.edu.cn.
\IEEEcompsocthanksitem Y. Shao is with the Department of Electrical and Electronic Engineering, Imperial College London, UK. 
E-mail: y.shao@imperial.ac.uk.}%
}

\IEEEtitleabstractindextext{%
\begin{abstract}
	This paper presents the first orthogonal frequency-division multiplexing(OFDM)-based digital over-the-air computation (AirComp) system for wireless federated edge learning, where multiple edge devices transmit model data simultaneously using non-orthogonal \rev{OFDM subcarriers}, and the edge server aggregates data directly from the superimposed signal. Existing analog AirComp systems often assume perfect phase alignment via channel precoding and utilize uncoded analog transmission for model aggregation. In contrast, our digital AirComp system leverages digital modulation and channel codes to overcome phase asynchrony, thereby achieving accurate model aggregation for phase-asynchronous multi-user OFDM systems. 
	To realize a digital AirComp system, we develop a medium access control (MAC) protocol that allows simultaneous transmissions from different users using non-orthogonal OFDM subcarriers, and put forth joint channel decoding and aggregation decoders tailored for convolutional and LDPC codes.
	To verify the proposed system design, we build a digital AirComp prototype on the USRP software-defined radio platform, and demonstrate a real-time LDPC-coded AirComp system with up to four users.
	Trace-driven simulation results on test accuracy versus SNR show that: 1) analog AirComp is sensitive to phase asynchrony in practical multi-user OFDM systems, and the test accuracy performance fails to improve even at high SNRs; 2) our digital AirComp system outperforms two analog AirComp systems at all SNRs, and approaches the optimal performance when SNR~$\geq$~6 dB for two-user LDPC-coded AirComp, demonstrating the advantage of digital AirComp in phase-asynchronous multi-user OFDM systems.
\end{abstract}

}

\maketitle
\IEEEdisplaynontitleabstractindextext
\IEEEpeerreviewmaketitle

\section{Introduction}\label{sec:introduction}

\IEEEPARstart{D}{ue} to the increased concern in data privacy-preserving, federated edge learning (FEEL) has become a popular distributed learning framework for mobile edge devices. A FEEL system often consists of an edge parameter server and multiple edge devices with non-independent and identically distributed (non-iid) local datasets. The parameter server periodically aggregates the locally-trained model parameters that are transmitted over the wireless channel.

\begin{figure}[t!]
	\centering
	\setlength{\abovecaptionskip}{0.cm}
	\includegraphics [scale=0.275]{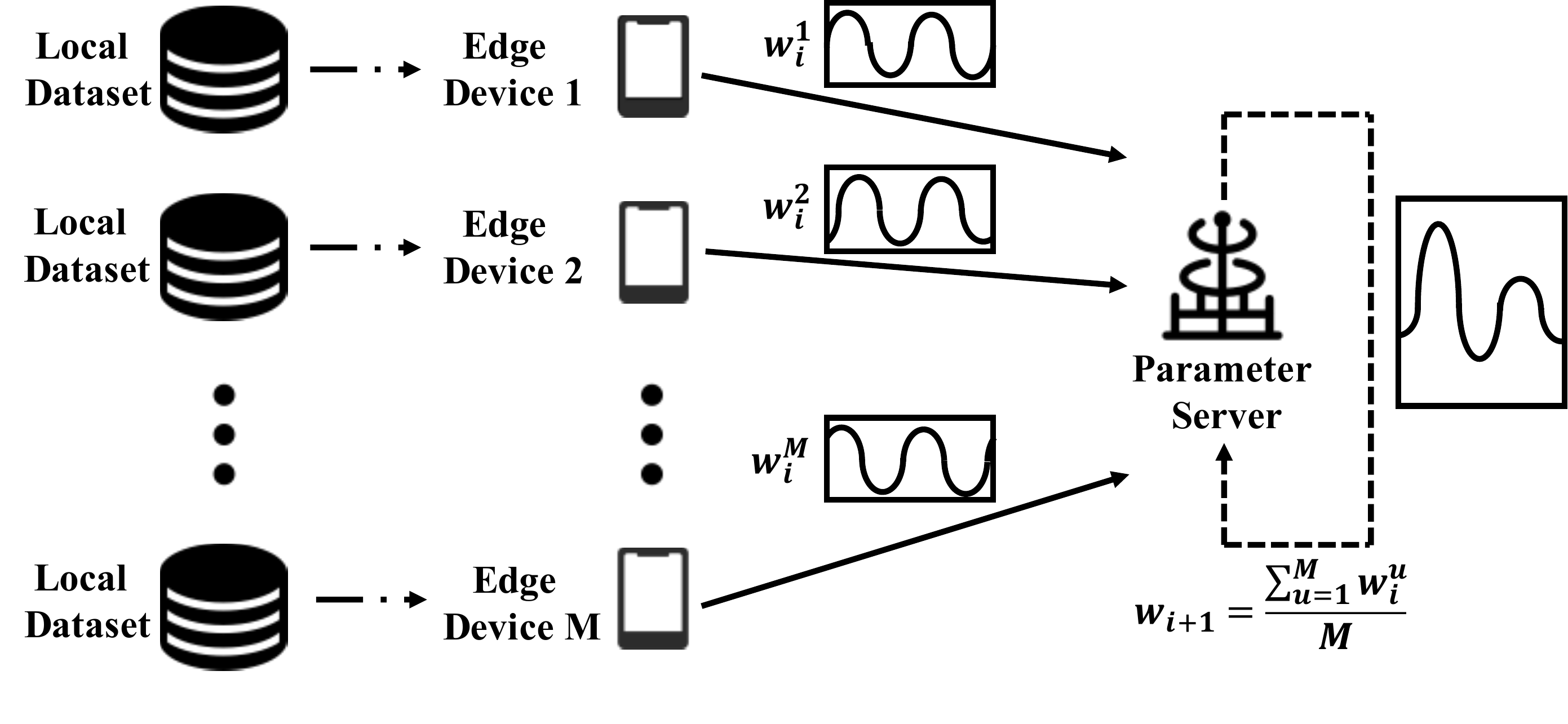}
	\caption{Model aggregation in wireless federated edge learning via over-the-air computation where edge devices transmit simultaneously and the average data is decoded directly from the superimposed signal.}\label{FL}
\end{figure}

Since large numbers of model parameters have to be transmitted in the model aggregation step, wireless communication becomes a bottleneck that restricts the practical development of FEEL \cite{McMahan2017}. Over-the-air computation (AirComp) has been proposed to address this problem\cite{amiri2019over,zhu2019broadband}, as shown in Fig.~\ref{FL}. In AirComp systems, selected edge devices transmit simultaneously using non-orthogonal frequency resources, and the parameter server decodes the arithmetic sum of edge devices’ data directly from the superimposed signal. \rev{The non-orthogonal channel access allows the access of more devices given the same frequency resource, and improves the communication efficiency. Decoding aggregation data directly at the physical layer avoids decoding individual data first and then aggregation, and may improve the computation performance.}

Existing AirComp systems \cite{amiri2019over, zhu2019broadband, xing2020decentralized, yang2020federated, amiri2020machine, zhu2020one, sery2021over, wang2021federated, shao2021denoising, wang2022interference, zou2022knowledge} are all analog AirComp systems that use discrete-time analog transmission for communication. Specifically, the edge devices transmit their model parameters in an analog fashion without channel coding. The overlapped signals in the air naturally produce the arithmetic sum. The parameter server then extracts the arithmetic sum directly from the superimposed signal. However, the performance of analog AirComp is sensitive to the phase offsets among the superimposed signal: they can add up destructively if their signals are asynchronous, resulting in large aggregation errors. Therefore, these analog AirComp systems assume perfect channel precoding that can compensate all channel impairments, such as path loss, shadowing, small-scale fading, time offset (TO), and carrier frequency offset (CFO), and receive the clean arithmetic sum of the transmission signals. 

However, realizing such perfect phase-aligned transmissions is quite challenging in practice, especially for random-access orthogonal frequency-division multiplexing (OFDM) systems such as WiFi. First, initial channel phase misalignment must be provided via a feedback protocol from the receiver, inducing some protocol overhead \cite{tan2018mobile}. Second, the residual CFO is unavoidable due to estimation errors. Even with perfect phase alignment initially, the residual CFO rotates subcarriers’ phases over time (symbols), and causes phase misalignment eventually.
For a multi-user OFDM system, different users may have different residual CFOs, leading to rotating phase offsets over symbols. As a consequence, the performance of analog AirComp deteriorates.

To tackle the problem, this paper puts forth the first OFDM-based digital AirComp system \rev{where multiple users use non-orthogonal subcarriers for data transmission}. Unlike analog AirComp that hinges on accurate channel precoding, digital AirComp leverages digital modulation and channel codes to combat channel impairments and phase misalignments, thereby achieving accurate model aggregation even when the signal phases of multiple edge devices are misaligned at the parameter server. Specifically, in channel-coded digital AirComp, model parameters are first quantized into bits, and then transmitted by a conventional OFDM encoder. Their signals get superimposed in the air.
Then, the receiver aims to decode the aggregated source bits (which we refer to as \emph{SUM bits}) instead of the individual source bits transmitted from different edge devices. 
To obtain SUM bits, the receiver first demodulates phase-asynchronous symbols, and then removes the protection of channel codes to get aggregation results. The latter process is referred to as joint channel decoding and aggregation (Jt-CDA).
\rev{Similar to the analysis in wireless quantized federated learning \cite{bouzinis2022wireless}, we prove that FEEL with digital AirComp can converge despite of SUM bit errors and quantization errors.}

Convolutional codes and LDPC codes are two common codes used in OFDM systems. We target supporting them in our digital AirComp system. However, it is challenging to implement low-complexity and real-time Jt-CDA decoders. For convolutional-coded AirComp, the SUM packet(codeword)-optimal decoder has prohibitively high computational complexity, since several codewords (and corresponding source bits) may lead to the same SUM bits, and their probability must be aggregated first before decoding. 
Thus, we design two reduced-complexity Jt-CDA decoders for convolutional-coded AirComp. The first Jt-CDA decoder, \emph{full-state joint decoder (FSJD)}, is a nearly optimal decoder that leverages the log-max approximation to realize maximum likelihood (ML) decoding. The second Jt-CDA decoder, \emph{reduced-state joint decoder (RSJD)}, is a simplified version of FSJD that greatly reduces the number of states without compromising much performance for most SNRs.

For LDPC-coded AirComp, we present a new belief propagation algorithm that realizes the SUM bit-optimal decoder. We reuse the Tanner graph and the decoding framework of message passing for conventional single-user LDPC codes, but redefine the passing message as a vector and the corresponding message update equations. 
The vector represents the probability of all users' bit combinations, and the update equations sum the probability of combinations that generate the same output.
After fixed rounds of iterations, we sum the probability of the bit combination that generates the same SUM bit, and choose the SUM bit with the maximal sum probability. The iterative BP algorithm and the per-bit mapping have low complexity and render it possible for real-time implementation.

We prototype these Jt-CDA decoders for digital AirComp on the USRP software-defined radio platform \cite{usrp} with the GNU Radio software \cite{gnuradio}, and perform extensive simulations and experiments to evaluate their performance against analog AirComp in phase-asynchronous FEEL systems. 
Simulation results show that phase offsets have a strong impact on the Jt-CDA performance.
Trace-driven simulation results on test accuracy verify that the analog AirComp fails to improve the performance in the presence of phase misalignments even at a high signal-to-noise ratio (SNR) regime. In contrast, digital AirComp completely avoids the problem, and outperforms analog AirComp at all SNRs. The proposed digital AirComp can approach optimal learning performance at high SNRs. Furthermore, we implement a real-time LDPC-coded AirComp system with up to four simultaneous transmissions. %
The real-time system leverages the graphic processing unit (GPU) hardware for highly parallel signal processing, and achieves a network throughput of 1.16 Mbps for four-user LDPC-coded AirComp considering the protocol overhead.

\rev{
The main technical contributions of this paper are summarized as follows:

\begin{enumerate}
    \item We present the first digital AirComp system that exploits the joint design of channel decoding and aggregation to overcome phase asynchrony, and prove the convergence of FEEL with digital AirComp. %
    \item We put forth three low-complexity Jt-CDA decoders for digital AirComp:  two decoders (i.e., FSJD and RSJD) for convolutional-coded AirComp and one decoder for LDPC-coded AirComp. %
    \item We study the physical-layer performance and the learning performance of FEEL with digital AirComp using experiment traces, and provide a real-time implementation of the LDPC-coded AirComp system.
\end{enumerate}
}

The remainder of this paper is organized as follows. Section \ref{sec:motivation} presents the background and the motivation for digital AirComp. Section \ref{sec:overview} overviews the proposed system.
Section \ref{sec:conv} and Section \ref{sec:ldpc} introduce the decoders for convolutional codes and LDPC codes. Experiment results are shown in Section \ref{sec:results}. 
\rev{Section \ref{sec:discussion} provides some discussions.}
Section \ref{sec:related} gives the related work. Section \ref{sec:conclusion} concludes the paper.

\section{Background and Motivation} \label{sec:motivation}

In this section, we first review the federated edge learning model, and the multi-user OFDM model without channel precoding. Then, we discuss the difficulties in realizing phase-aligned transmissions via channel precoding, and advocate digital AirComp. Finally, we show the convergence and benefit of federated learning under digital AirComp.

\subsection{Wireless Federated Edge Learning}
In FEEL, multiple edge devices \rev{(e.g., $Q$ devices)} collaboratively learn a joint model with the help of a parameter server without transmitting raw datasets to each other. 
The federated model training requires many iterations to converge. Each iteration contains the following four steps:
\begin{enumerate}
\item Model broadcast: \rev{$P$} edge devices are randomly selected by the parameter server and each of them is broadcasted with the current global model $w_t$ in the $t$-th iteration;
\item Local training: every edge device $u$ trains the received global model with their local data and gets a new local model $w_t^u$;
\item Model update: the local models of selected edge devices are transmitted to the parameter server;
\item Model aggregation: the parameter server takes the average of the received models to update the global model for the next $t$+1-th iteration: $w_{t+1}=\sum_{u=1}^{\rev{P}}w^u_t/\rev{P}$.
\end{enumerate}

For mobile edge devices, FEEL is often realized by wireless networks (e.g., WiFi, 5G). In the following, we focus on step 3) and 4), and study how to apply over-the-air computation to improve the model aggregation performance.

\subsection{Multi-User OFDM System Model} \label{sec:motivation:ofdm}

The model update and aggregation steps require a large bandwidth, since each edge device needs to transmit a large number of model coefficients to the parameter server.
Therefore, in this paper, we consider OFDM wideband systems for the model update, especially WiFi OFDM systems. %
Different from traditional OFDM systems, AirComp uses non-orthogonal channel access \rev{where all edge devices transmit data at the same OFDM subcarriers}.

For user $u$, our system first quantizes the updated model parameters with a quantization length of $b$ bits per parameter. Then we pack several parameters in a packet for transmission. Assuming an OFDM packet can transmit $n$ source bits, we can pack $\lceil n / b \rceil$ parameters in a packet. 
Let $S^u=\{s^u_0, \dots, s^u_{n-1} \}$ be the source bits fed to the OFDM transmitter. After channel encoding, the coded bits $C^u$ are modulated in the frequency domain. We assume that binary phase shift keying (BPSK) is used in this paper, and each bit $C[l]$ is modulated to $X^u[l]$ with $X^u[l] \in \{+1,-1\}$. Then after inverse discrete Fourier transform (IDFT), cyclic prefix, preambles padding, and digital-to-analog conversion, $X^u$ is transformed into a time-domain continuous signal $x^u(t)$.

Let $\mathcal{U}$ be the set of users that transmit simultaneously. %
Let $\tau^u$ be the time offset (TO) with respect to the start of receiving window at the parameter server, and $\delta^u$ be the carrier frequency offset (CFO) between user $u$ and the receiver.
Then, the received baseband signal $y(t)$ can be represented as
\begin{equation} \label{eqn:time}
	y(t) = \sum_{u \in \mathcal{U}} (h^u(t) * x^u(t-\tau^u)) e^{j 2 \pi \delta^u t} + n(t),
	\vspace{-1ex}
\end{equation}
where $h^u(t)$ is the impulse response of the channel between user $u$ and the receiver, $*$ denotes the convolution operation, and $n(t)$ is the zero-mean circularly-symmetric complex additive white Gaussian noise (AWGN). \rev{Here $h^u(t)$ includes large-scale and small-scale fading, and thus is a complex vector under multipath channel. Let $h^u (t)=[h_1^u,\dots,h_L^u]$ where there are $L$ paths and the $l$-th path has delay $\tau_l^u$.}

Most commercial multi-user OFDM systems (e.g., WiFi 6, 5G) specify the maximal time synchronization error and frequency synchronization error so that TO and CFO do not affect the decoding performance. For example, WiFi 6 enforces a timing accuracy of less than $\pm$0.4$\mu$s and a CFO error less than 350Hz (i.e., 0.07 ppm at 5GHz) after the trigger frame for multi-user orthogonal transmissions \cite{wifi6}. Although the accuracy is specified, the TO and CFO of users are very likely to be different in practical systems, i.e., $\tau^{u_1} \neq \tau^{u_2}$ and $\delta^{u_1} \neq \delta^{u_2}$ ($u_1,u_2 \in \mathcal{U}$).

Given such accuracy, the inter-subcarrier interference (ICI) is negligible, and the receiver can find a proper cyclic prefix cut without affecting the decoding performance. However, TO introduces phase rotation in each subcarrier, and CFO also causes phase rotation in each subcarrier over symbols. Specifically, the received baseband signal of the $i$-th symbol and the $k$-th subcarrier in the frequency domain can be represented as
\begin{equation}  \label{eqn:freq}
	Y_i[k]= \sum_{u \in \mathcal{U}}  H^u[k] X_i^u[k] e^{j 2 \pi \frac{\tau^{u} k}{\rev{N_s}}} e^{j 2 \pi \delta^u i (\rev{N_s+N_{cp}})} +N_i[k],
\end{equation}
where $H^u$ is the fast Fourier transform (FFT) of $h^u$, $X_i^u$ is the modulated signal in the frequency domain, \rev{$N_s$} is the number of FFT points, and \rev{$N_{cp}$} is the number of cyclic prefix (CP) samples.
\rev{
Assume the demodulation window is aligned with the first path. Then we have
\begin{equation}
    H^u[k]=\sum_{l=1}^L h_l^u e^{j\frac{2\pi\tau_l^u k}{N_s}}.
\end{equation}
According to the above formula, the multi-path channel introduces phase misalignment even with the CP design in OFDM since $h_l^u$ and $\tau_l^u$ are different from different users.
}

Note that TO causes a linear phase rotation across subcarriers. On the other hand, although the inter-subcarrier interference (ICI) caused by a CFO value of 350Hz is negligible, CFO still causes constant phase rotation across subcarriers, and phase rotation accumulates over symbols.
Since TO does not cause phase accumulation over symbols, Eq.~(\ref{eqn:freq}) can be simplified as
\begin{equation}  \label{eqn:freq2}
	Y_i[k]= \sum_{u \in \mathcal{U}}  A^u[k] X_i^u[k] e^{j 2 \pi \delta^u i (\rev{N_s+N_{cp}})} +N_i[k],
\end{equation}
where $A^u[k] \triangleq H^u[k] e^{j 2 \pi \frac{\tau^{u} k}{\rev{N_s}}}$.

Our target is to compute the arithmetic sum $S^F \triangleq \sum_{u \in \mathcal{U}}S^u=\{\sum_{u \in \mathcal{U}}s_0^u, \cdots, \sum_{u \in \mathcal{U}}s_{n-1}^u\}$ of all parameters from $Y$, where operator in $\sum$ is the arithmetic sum. Then, we can easily compute the average.

\subsection{Challenges of Phase-Aligned Transmissions} \label{sec:motivation:misaligned}

\begin{figure}[t!]
	\centering
	\setlength{\abovecaptionskip}{0.cm}
	\includegraphics [scale=0.35]{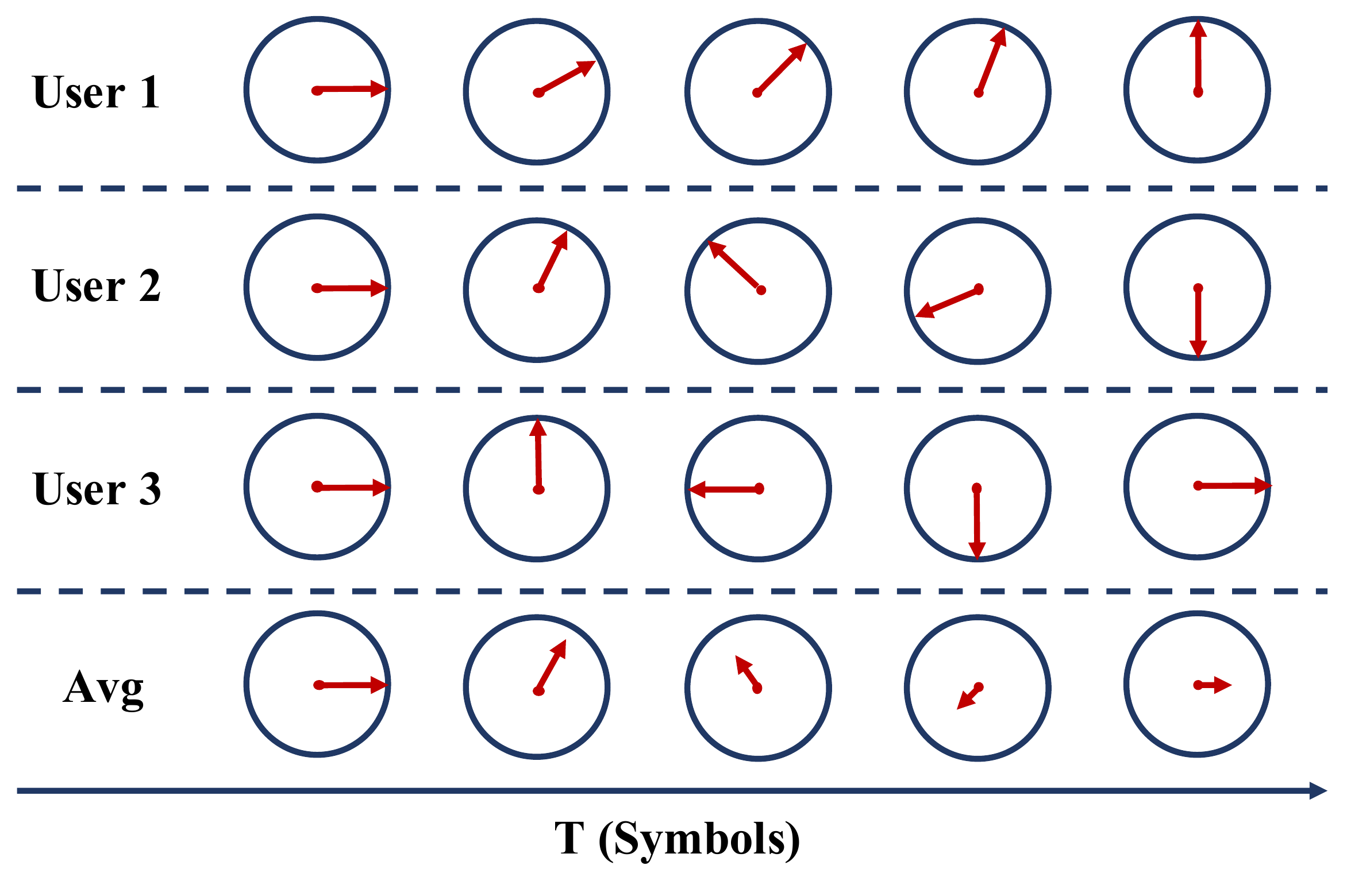}
	\caption{The average operation of an analog AirComp system with imperfect channel precoding in one OFDM subcarrier over  symbols.}\label{fig:analog_fedavg_error}
	\vspace{-1ex}
\end{figure}

Existing OFDM-based AirComp systems \cite{zhu2019broadband, zhu2020one} assume near-perfect phase alignment using channel precoding. 
Such channel precoding is implemented in an analog AirComp system recently \cite{guo2021over}. Multi-user OFDM systems with channel precoding have been demonstrated in network MIMO systems \cite{rahul2012jmb,balan2013airsync, abari2015airshare, wang2017dcap} and a physical-layer network coding (PNC) system \cite{tan2018mobile}. 
Nevertheless, considerable protocol overhead is required for the initial channel phase alignment from different users. 
Even if the channel phases are aligned initially, maintaining such a phase alignment is quite costly due to residual CFO after precoding, especially when the packet duration is large.

In Eq.~(\ref{eqn:freq2}), due to the existence of $A^u[k]$ and $\delta^u$, their channels are not phase synchronous. The task of a channel precoding protocol is to eliminate the power difference and phase difference caused by $H^u[k]$, $\tau^u$ and $\delta^u$. In particular, the receiver estimates composed channel $\hat{A}^u[k]$ and  CFO $\delta^u$, and sends them back to the transmitter. The transmitter precodes its frequency-domain signal $S_i^u[k]$ with a coefficient $P_i^u[k]$ in the $k$-th subcarrier of the $i$-th symbol (i.e., $X^u_i[k]=P^u_i[k]S^u_i[k]$), where 
$$P^u_i[k]=\frac{e^{-j 2 \pi \hat{\delta}^u i(\rev{N_s+N_{cp}})}}{\hat{A}^u[k]}.$$
Then the received signal with channel precoding is represented by
\begin{equation}  \label{eqn:precode-freq2}
	Y_i[k]= \sum_{u \in \mathcal{U}}  \Delta A^u[k] S^u_i[k] e^{j 2 \pi \Delta\delta^u i(\rev{N_s+N_{cp}})} +N_i[k],
\end{equation}
where $\Delta A^u[k]=\frac{A^u[k]}{\hat{A}^u[k]}$ and $\Delta\delta^u=\delta^u-\hat{\delta}^u$.

In practical systems with channel precoding, the receiver needs to estimate $\hat{A}^u[k]$ on the time granularity of the channel coherence time, and quantize $\hat{A}^u[k]$ for all subcarriers for feedback. Moreover, $\Delta A^u[k] \neq 1$ and $\Delta\delta^u \neq 0$ due to estimation errors. The phase offset caused by $\Delta A^u[k]$ is constant over the whole packet, and thus it does not affect decoding with accurate precoding. However, the phase offset caused by $\Delta\delta^u$ accumulates as time goes, and can easily lead to severe phase misalignment for a long packet. Fig.~\ref{fig:analog_fedavg_error} illustrates such a phenomenon and its impact on the average operation of analog AirComp in one OFDM subcarrier. As we can see, the phase misalignment over symbols degrades the performance of analog AirComp.

Highly-accurate CFO precoding is not easy to achieve in practical systems. For example, given the $\pm$0.07ppm frequency error specified by WiFi 6 (assuming the central frequency 5GHz), 1.5ms will lead to a phase offset of around $2 \times 0.07\cdot10^{-6} \times 5\cdot10^9 \times  1.5\cdot10^{-3} \times 2 \pi \approx \pi$ radian (i.e., totally misaligned) in the worst case. 
Assuming that the phase offset must be less than $\pi$/4 radian for constructive signal superimposition, a packet must be no more than 100 symbols (the duration of each symbol is 4$\mu$s for 20MHz bandwidth and 80 samples per symbol). The packet duration corresponds to 300Bytes (assuming BPSK and 1/2 coding rate), which is not adequate for FEEL.

To tackle the problem, many approaches have been proposed. A high-precision but expensive clock can be used (e.g., global position system disciplined oscillator clock with 0.1ppb \cite{jackson}). Protocols with some signaling overhead are used in the two-way relay network \cite{tan2018mobile} and the multiple access network \cite{guo2021over, csahin2022demonstration}. Extra radio \cite{balan2013airsync, abari2015airshare} or extra infrastructure \cite{rahul2012jmb, yenamandra2014vidyut, wang2017dcap} are required for network MIMO.
\rev{In this paper, we leverage the digital AirComp technique to overcome the phase asynchrony problem.}

\rev{
\subsection{Convergence of Digital AirComp-enabled FEEL} \label{sec:motivation:convergence}
Although our digital AirComp system can decode aggregate data, some sum bits are inevitably erroneous. Given that most FEEL algorithms work under the error-free assumption, the first question is whether such sum bit errors affect the convergence of federated learning. 

For simplicity, we assume that the model broadcast step is error-free and only focus on finding out how the SUM bit error rate (SUM BER) of the received superimposed signal and quantization precision in the model aggregation step affect the convergence rate of our proposed system. The proof is similar to \cite{bouzinis2022wireless} that considers quantization errors, but differs in that we further consider the transmission errors in the uplink (i.e., the SUM BER).

We consider an FL process consisting of $T$ communication rounds, and each round contains $\tau$ steps of the stochastic gradient descent (SGD) operation. Without loss of generality, we assume all edge devices participate in the training, and
$M$ is the number of edge devices. We assume the each device transmits $d$ parameters of the local model in each communication round. The goal of the training process is to find the parameter $w^*$ that minimizes the loss function of the whole dataset
\begin{equation}
    F(w)=\sum_{n=1}^Mp_nF_n(w),
\end{equation}
where $p_n$ is the fraction of local datasets among total datasets and $F_n(w)$ is the loss function of edge device $n$. That is, $w^*=\mathop{\arg\min}_{w} F(w)$.

Let $w(t)$ be the broadcasted model at the beginning of the $t$-th communication round. Take the $n$-th edge device for example. The $i$-th step of SGD can be represented as
\begin{equation}
    w_n^i(t)=w_n^{i-1}(t)-\eta(n)\nabla F_n(w_n^{i-1}(t),\xi_n^{i-1}(t)),
\end{equation}
where $w^0_n(t)=w(t)$, $\eta(n)$ represents the learning rate, and $\xi_n^{i}(t)$ represents the batch size. Then the local weight difference in the $t$-th communication round is defined as
\begin{equation}
    \Delta w_n(t)=w_n^\tau(t)-w(t).
\end{equation}
We define the global model at the parameter server of the $t$-th communication round as
\begin{equation}
    w(t+1)=w(t)+\frac{1}{M}\sum_{n=1}^Mp_nB(\Delta w_n(t)),
\end{equation}
where $B(\Delta w_n(t))$ is the updated weight differential from selected edge device $n$.

The correctness of the global weight is determined by the quantization precision and SUM BER of the transmission, which can be represented as
\begin{equation}
    B(\Delta w_n(t))=Q(\Delta w_n(t))+X(\Delta w_n(t)),
\end{equation}
where $Q(\Delta w_n(t))$ is the updated weight error after quantization and $X(\Delta w_n(t))$ is the update weight error caused by communication.

To facilitate the convergence analysis, we use the single-user BER $\alpha$ for analysis in each communication round.
The exact relationship between BER $\alpha$ and SUM BER $\beta$ is shown in Appendix A.
Let $B_n(t)$ be the number of quantization bits. Assume $\mathcal{K}$ contains the erroneous positions where $\mathcal{K} \subset \{1, \cdots, B_n(t)\}$ in the $t$-th communication round. The communication error induced by $\mathcal{K}$ is given by 
\begin{equation}
    \begin{aligned}
        X^\mathcal{K}(\Delta w_n(t)) &= 
        \sum_{k\in \mathcal{K}} X^k(\Delta w_n(t)) \\
        &=\sum_{k\in \mathcal{K}} \frac{\Delta w_n^{max}(t)-\Delta w_n^{min}(t)}{2^{B_n(t)}-1}&\times 2^{k-1} \times m^k,
    \end{aligned}
\end{equation}
where $\Delta w_n^{max}(t)$ and $\Delta w_n^{min}(t)$ are the maximum and minimum value of weight difference vector after edge device $n$ performs local training, and $m^k$ is an indicator that is -1 if the error of the $k$-th bit is $1\rightarrow{0}$, or 1 if the error is $0\rightarrow{1}$. Note that $\mathbb{E}(X^\mathcal{K}(\Delta w_n(t)))=0$, since each bit has an equal probability for $0\rightarrow{1}$ and $1\rightarrow{0}$.

Let $\mathbb{K}$ be all possible error patterns. Then
\begin{equation} \label{22}
    \mathbb{E}[X(\Delta w_n(t))]=\sum_{\mathcal{K} \in \mathbb{K}} p^{\mathcal{K}} \mathbb{E}[X^\mathcal{K}(\Delta w_n(t))]=0,
\end{equation}
where $p^{\mathcal{K}}$ is the probability that bits in $\mathcal{K}$ are all erroneous. 

Then, we have the following lemma.
\newtheorem{lemma}{Lemma}
\begin{lemma}\label{lemma1}
$B(\Delta w_n(t))$ is an unbiased estimator of $\Delta w_n(t)$, i.e.,
\begin{equation}
\mathbb{E}[B(\Delta w_n(t))] = \Delta w_n(t),
\end{equation}
while it also holds that
\begin{equation} \label{22}
    \begin{aligned}
        &\mathbb{E}[\Vert B(\Delta w_n(t))-\Delta w_n(t)\Vert_2^2] \leq J_n^2(t)+K_n(t),
    \end{aligned}
\end{equation}
where 
\begin{equation}
    \begin{aligned}
        & J_n^2(t) \triangleq \frac{\delta_n^2(t)}{(2^{B_n(t)}-1)^2}, \\
        & K_n(t) \triangleq \alpha \left(\frac{4\delta_n(t)^2}{d(2^{B_n(t)}-1)^2}\right)^2 \times \frac{1-4^{B_n^2(t)}}{1-4}, \\
        & \delta_n(t) \triangleq \sqrt{\frac{d}{4}(\Delta w_n^{max}(t)-\Delta w_n^{min}(t))^2}. \nonumber
    \end{aligned}
\end{equation}
\end{lemma}
\begin{proof}
See Appendix B.
\end{proof}

Next, we use the following commonly used assumptions to prove convergence.

\newtheorem{assumption}{Assumption}
\begin{assumption}
    ($\mu$-strongly convex). For each edge device $n$, function $F_n$ is $\mu$-strongly convex, which means, for all $w$ and $w'$, we have $F_n(w') \geq F_n(w) + \langle w'-w,\nabla F_n(w)\rangle +\frac{\mu}{2}\vert w'-w \Vert_2^2$.
\end{assumption}
\begin{assumption}
    (L-smoothness). For each edge device $n$, function $F_n$ is L-smooth, which means, for all $w$ and $w'$, we have $F_n(w') \leq F_n(w) + \langle w'-w,\nabla F_n(w)\rangle +\frac{L}{2}\vert w'-w \Vert_2^2$.
\end{assumption}
\begin{assumption}    
    (Uniformly bounded). For each edge device $n$, communication round $t$, and local SGD step $i$, the expected squared norm of local stochastic weight differences is uniformly bounded, $\mathbb{E}[\Vert \nabla F_n(w_n^i(t),\xi_n^i(t))\Vert_2^2]\leq G^2$.
\end{assumption}
\begin{assumption}
    (Variance bounded). For each edge device $n$, communication round $t$, and local SGD step $i$, the estimate of local stochastic gradients has a bounded variance, $\mathbb{E}[\Vert \nabla F_n(w_n^i(t),\xi_n^i(t)-\nabla F_n(w_n^i(t))\Vert_2^2]\leq \sigma_n^2$.
\end{assumption}

We define $\Gamma$ as the degree of non-iid among user's datasets, $\Gamma\triangleq F(w^*)-\sum_{n=1}^{M}p(n)F_n^*,$ where $F_n^*$ represents the minimum value of $F_n$. With the above equations and assumptions, we can have the following theorem.

\newtheorem{tho}{Theorem}
\begin{tho}\label{theorem1}
    By selecting a diminishing learning rate $\eta(t)=\frac{2}{\mu(\gamma+t)}$ and $\gamma > max\left\{2,\frac{2}{\mu},\frac{L}{\mu}\right\}$, $\mathbb{E}[F(w(T)-F(w^*))]$ is upper bounded by
    \begin{equation}
        \small
        \begin{aligned}
            \mathbb{E}&[F(w(T))-F(w^*)]\leq \\
            &\frac{L}{2}\frac{1}{\gamma+T}(\frac{4U}{\mu^2}+\gamma \mathbb{E}[\Vert w(0)-w^*\Vert_2^2]) \\
            &+\frac{L}{2}\sum_{j=0}^{T-1}\left[ \sum_{n=1}^Mp_n(J_n^2(t)+K_n(t))\prod_{i=j+1}^{T-1}(1-\frac{2}{\gamma+i}) \right],
        \end{aligned}
        \label{upper_bound}
    \end{equation}
    where 
    \begin{equation}
        \small
        \begin{aligned}
        U=\tau^2\sum\sigma_n^2+\tau G^2+2L\tau^2\Gamma+(\mu+2)\frac{\tau(\tau-1)(2\tau-1)}{6}G^2 \nonumber.
        \end{aligned}
    \end{equation}
\end{tho}
\begin{proof}
See Appendix B.
\end{proof}

In Eq.~(\ref{upper_bound}), the first term of the upper bound tends to zero for large $T$. For the second term, compared with results in \cite{bouzinis2022wireless}, the added element $K_n(t)$ is related to BER $\alpha$. The second term related to quantization error and transmission error indeed causes a gap. However, following the same argument in \cite{bouzinis2022wireless}, the term $\prod_{i=j+1}^{T-1}(1-\frac{2}{\gamma+i})$ tends to zero for small $j$, which means the gap may be negligible in practice.

}

\begin{figure}[t]
	\centering
	\subfloat[ShuffleNet-V2 with CIFAR-10]{  %
		\includegraphics[width=0.46\columnwidth]{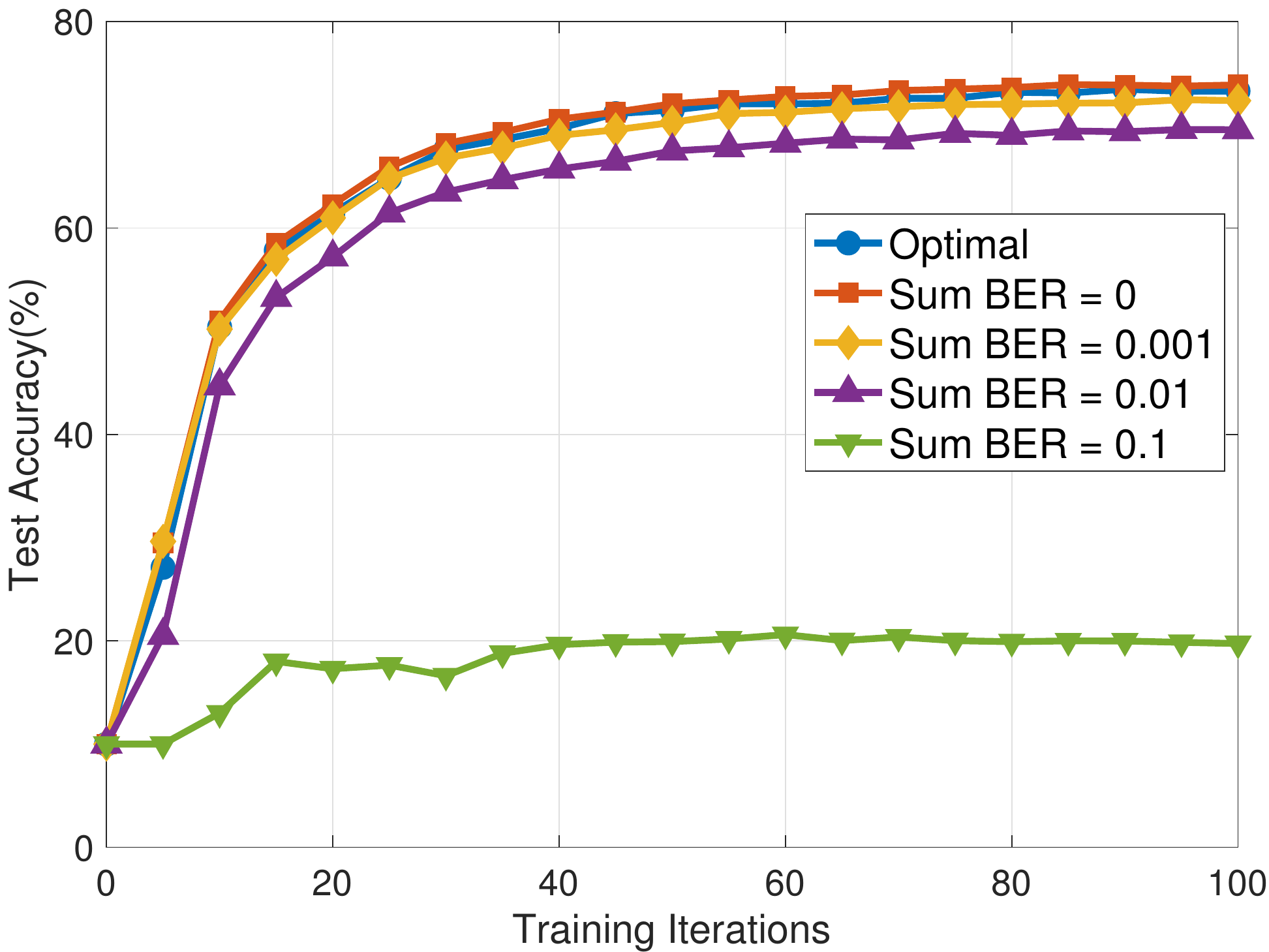}
        \label{shuffle}
    }
    \hspace{1mm}
	\subfloat[MLP with MNIST]{
		\includegraphics[width=0.46\columnwidth]{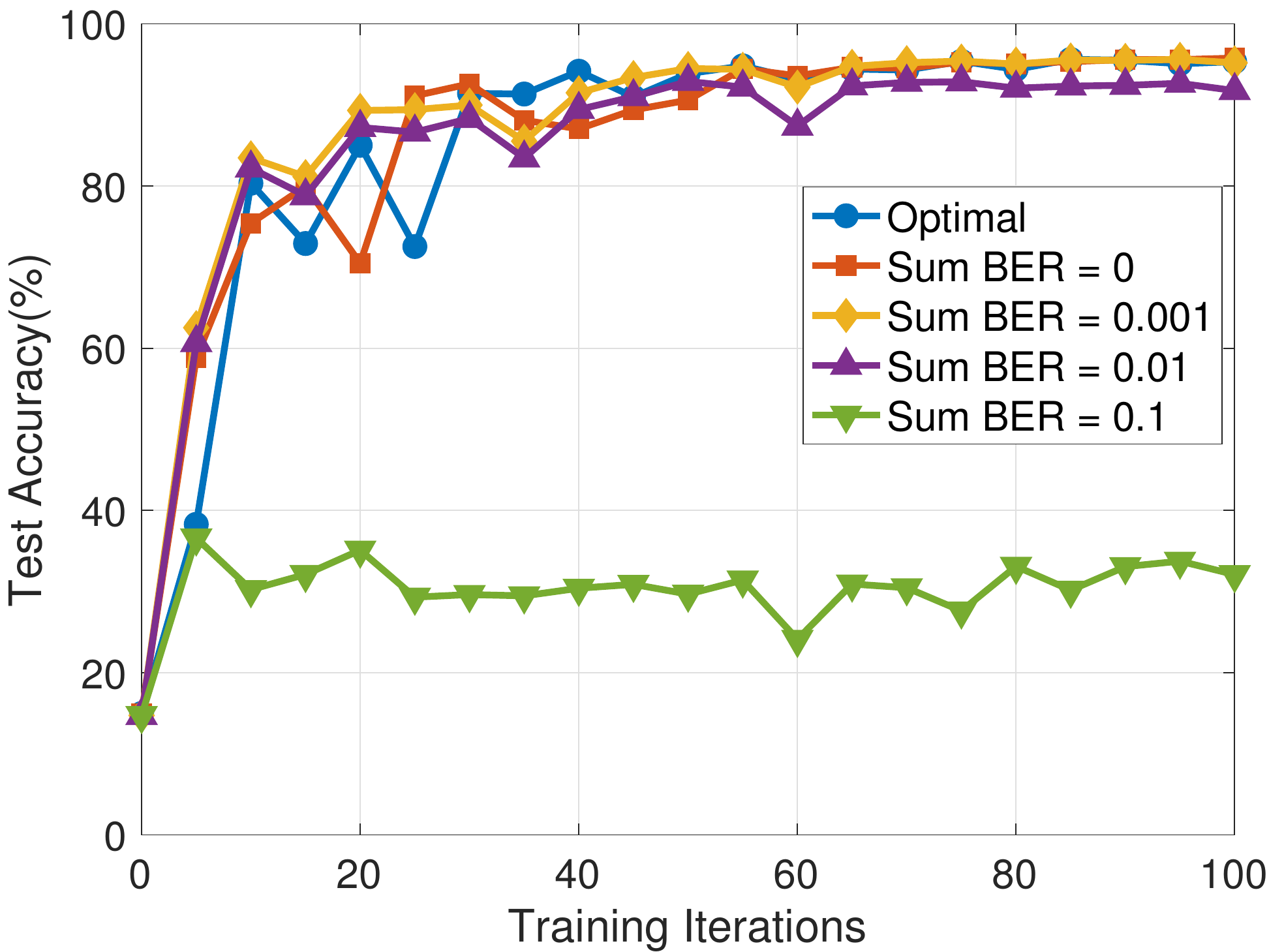}
        \label{mlp}
    }
	\caption{Test Accuracy Performance of two digital AirComp-enabled FEEL systems under different SUM BERs.}
	\label{fl}
\end{figure}

\rev{\textbf{Simulation Results.}
We further study the convergence of the digital AirComp-enabled FEEL system, and the impact of SUM BER on the learning accuracy of FEEL applications by simulation.} We choose two typical FL applications with non-iid data (the data allocation method can be found in Section~\ref{sec:eval:fl}): CIFAR-10 with ShuffleNet-V2 \cite{Krizhevsky2009} and MNIST with multi-layer perceptron (MLP) \cite{deng2012mnist}, and investigate their performance under different SUM BERs.
We adopt 8-bit probability quantization  \cite{suresh2017distributed} for each model parameter.
There are 40 edge devices in total, and 4 devices are randomly chosen for training each time. The test accuracy results are shown in Fig.~\ref{fl}(a) and Fig.~\ref{fl}(b), where the optimal curve shows the result when model parameters are not quantized and aggregated without error. They can converge to the optimal if SUM BER is smaller than $10^{-3}$, and converge close to the optimal if smaller than $10^{-2}$.

\section{System Overview} \label{sec:overview}

\begin{figure}[t!]
	\centering
	\hspace{-2ex}
	\setlength{\abovecaptionskip}{2.mm}
	\includegraphics [scale=0.325,trim=0 0 0 0]{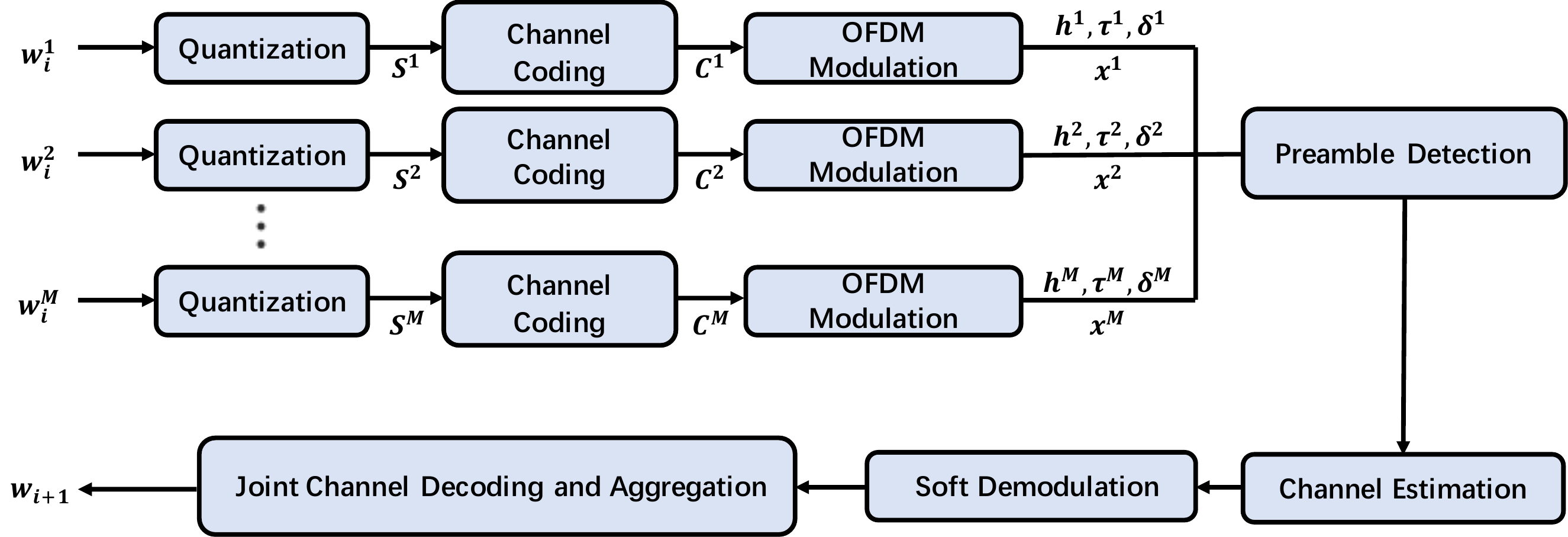}
	\caption{The architecture of the OFDM-based digital AirComp transceiver.}\label{fig:arch}
\end{figure}

\rev{There are $Q$ devices in the network. In each FEEL iteration, $P$ edge devices are chosen to participate in model update. They use the digital AirComp system that allows $M$ devices to transmit simultaneously, and require $\lceil P/M \rceil$ rounds of transmissions in each iteration. In the following, we focus on one round of the digital AirComp transmission.}

\textbf{Architecture.}
Fig.~\ref{fig:arch} shows the transceiver architecture of our digital AirComp system and the workflow. To be best compatible with the WiFi standards, our system reuses most components as in conventional WiFi OFDM systems and makes minimal modifications. In particular, the transmitter architecture is almost the same as the traditional OFDM transmitter. %
For the receiver, we follow the conventional frame detection and multi-user channel estimation design, but redesign the demodulation and channel decoding for the arithmetic sum computation. Demodulation is a soft demodulation block where the probabilities of all possible constellation combinations are preserved. 

Channel decoding and aggregation are performed jointly, which is referred to as \emph{Jt-CDA}. The Jt-CDA decoder is the newly designed block in digital AirComp. Given that both convolutional codes and LDPC codes are mandatory in WiFi 6 802.11ax \cite{wifi6}, we aim to support these two codes. 
The design for convolutional codes is presented in Section~\ref{sec:conv}, and the design for LDPC codes is presented in Section~\ref{sec:ldpc}.

\begin{figure}[t!]
	\centering
	\setlength{\abovecaptionskip}{0.cm}
	\includegraphics [scale=0.265,trim=0 0 0 0]{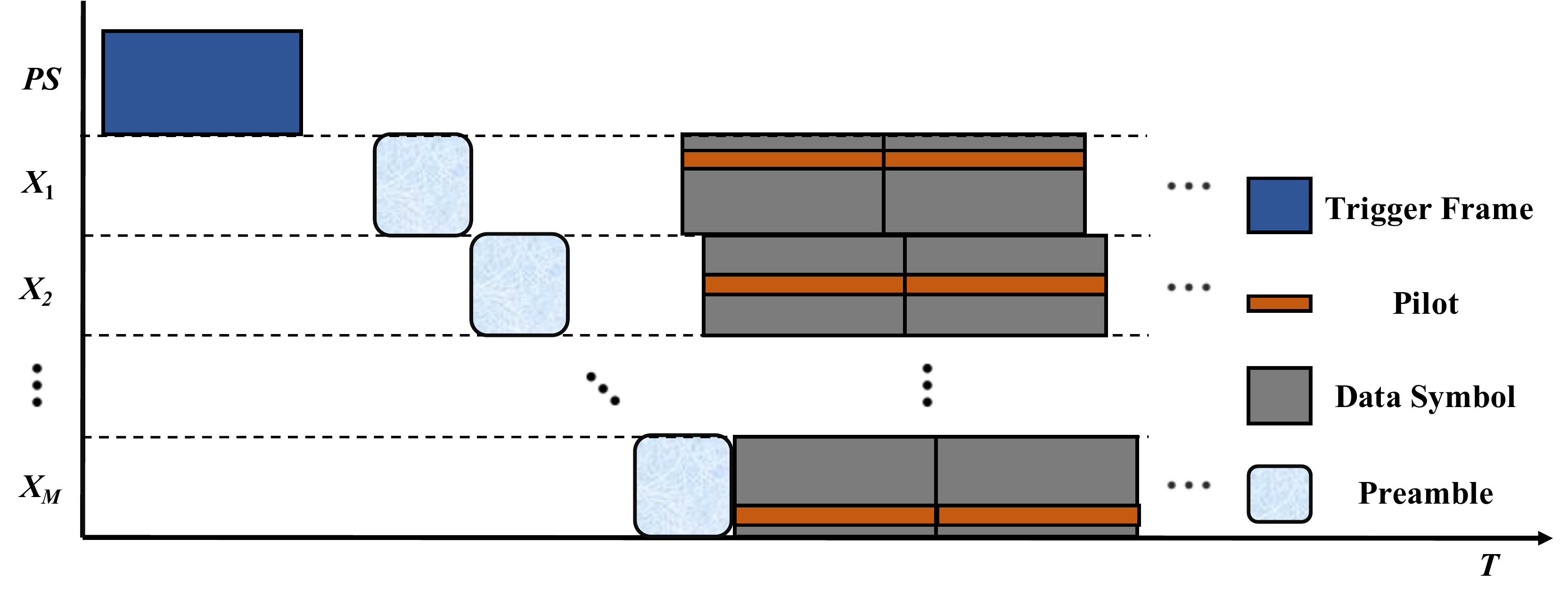}
	\caption{A MAC protocol for digital AirComp where the parameter server contends the channel on behalf of all users following the traditional random access protocol in Wi-Fi networks.}
	\label{nnma}
\end{figure}

\textbf{Protocol.}
Fig.~\ref{nnma} shows the enabling medium access control (MAC) protocol for multi-user non-orthogonal channel access.
Similar to the multi-user simultaneous transmission protocol in 802.11ax (e.g., UL-MIMO), the parameter server contends the channel on behalf of all users following the traditional random access protocol in Wi-Fi networks, and then transmits a trigger frame to notify the transmission users in each round. \rev{The trigger frame includes the coding scheme information (e.g., which channel codes and coding rate to use) and the power control information (e.g., the used transmission power and the expected reception power).} All informed users synchronize to the trigger frame and then transmit OFDM frames simultaneously with their preambles orthogonal in time domain, pilots orthogonal in frequency domain, and data symbols overlapped in time domain.  \rev{Furthermore, all informed users use the provided power information to perform fine-grained power control to ensure the reception powers of their signals are near-balanced. Such a power control scheme is valid due to the existence of channel reciprocity where the uplink channel and the downlink channel go through almost the same attenuation in the same frequency, and has also been adopted for multi-user simultaneous transmissions (e.g., OFDMA and UL MU-MIMO) in IEEE 802.11ax \cite{wifi6}. %
}

Each OFDM frame consists of 10 short training sequences (STS) and 2 long training sequences (LTS), and some data symbols.
Moreover, we allocate orthogonal pilots to each edge device in all OFDM data symbols. In this way, the receiver can estimate the initial frequency-domain channel through orthogonal LTSes, and track phase rotation through orthogonal pilots. Then the receiver performs Jt-CDA to compute average model parameter coefficients. In particular, for the $t$-th round, device $u$ quantizes its weight $w_t^u$ to $S_t^u$, encodes and modulates $S_t^u$ to $x_t^u(t)$. The received signal is shown as Eq.~(\ref{eqn:time}) in time domain and Eq.~(\ref{eqn:freq}) in frequency domain. In the end, the receiver extracts the average weight $w_{t+1}$ for the next round.

\section{Design for Convolutional Codes} \label{sec:conv} 
Convolutional codes are widely used in wireless communication systems (e.g., WiFi, Cellular, Satellite).%
For single-user communication systems, convolutional codes are often decoded by the Viterbi algorithm that performs maximum likelihood (ML) decoding and achieves the codeword-optimal performance. However, for multi-user simultaneous transmission systems, such SUM codeword-optimal performance is not easy to achieve.

\subsection{SUM Packet-optimal Decoder}
For ease of illustration, we assume there are only two concurrent users $A$ and $B$. The extension to more users is straightforward.
Their source packets are $S^A$ and $S^B$ with length $K$. Let the encoding function be $\Pi(\cdot)$. Their encoded packets (codewords) are $C^A=\Pi(S^A)$ and $C^B=\Pi(S^B)$ with length $N$. Their modulated signals are denoted by $X^A$ and $X^B$. Let $Y$ denote the received signal. Our target is to get the arithmetic sum of source packets $S^F = S^A + S^B$. Note that there is no valid codeword corresponding to $S^F$ since the arithmetic sum $+$ operation is not 
closed under $\mathcal{F}_2$ and $\Pi(\cdot)$. Therefore, we also call it \emph{SUM packet-optimal decoder}.
More specifically, our target is to compute 

\begin{small}
\begin{equation} \label{eq1}
	\vspace{-1ex}
	\begin{split}
		\hat{S}^{F} &= \operatorname*{arg\,max}_{S^{F}} \sum_{C^A, C^B: \Pi^{-1}(C^A)+\Pi^{-1}(C^B)=S^F} Pr(Y|C^A,C^B) \\
		&= \operatorname*{arg\,max}_{S^{F}} {log} \sum_{C^A, C^B: \Pi^{-1}(C^A)+\Pi^{-1}(C^B)=S^F} Pr(Y|C^A,C^B),
	\end{split}
	\vspace{-1ex}
\end{equation}
\end{small}
where
\begin{small}
\begin{equation} \label{eq2}
	\vspace{-1ex}
	\begin{split}
		&Pr(Y|C^A,C^B)=\prod_{n=1}^N Pr(Y[n]|C^A[n],C^B[n]) \\
		&\propto \prod_{n=1}^N exp(-\frac{\Vert Y[n]-H^A[n]X^A[n]-H^B[n]X^B[n] \Vert^2}{2\sigma^2}) \\
		&\propto exp(\sum_{n=1}^N -\frac{\Vert Y[n]-H^A[n]X^A[n]-H^B[n]X^B[n] \Vert^2}{2\sigma^2}).
	\end{split}
	\vspace{-1ex}
\end{equation}
\end{small}

Eq.~(\ref{eq1}) means that we need to add up the probability of all possible codeword pairs ($C^A$, $C^B$), aggregate the probability of pairs that yield the same $S^F$, and then make decision. The functional mapping from $S^A$ and $S^B$ to the SUM packet $S^F$ can be expressed as
$$f_{packet}: \{0, 1\}^K \times \{0, 1\}^K \rightarrow \{0, 1, 2\}^K.$$
In fact, it is a $(4/3)^K$-to-1 mapping. There is no known exact computation method for Eq.~(\ref{eq1}) except for exhaustively sum over all possible combinations, which has prohibitively high computational complexity.

\begin{figure}[t!]
	\centering
	\setlength{\abovecaptionskip}{0.cm}
	\includegraphics [scale=0.275]{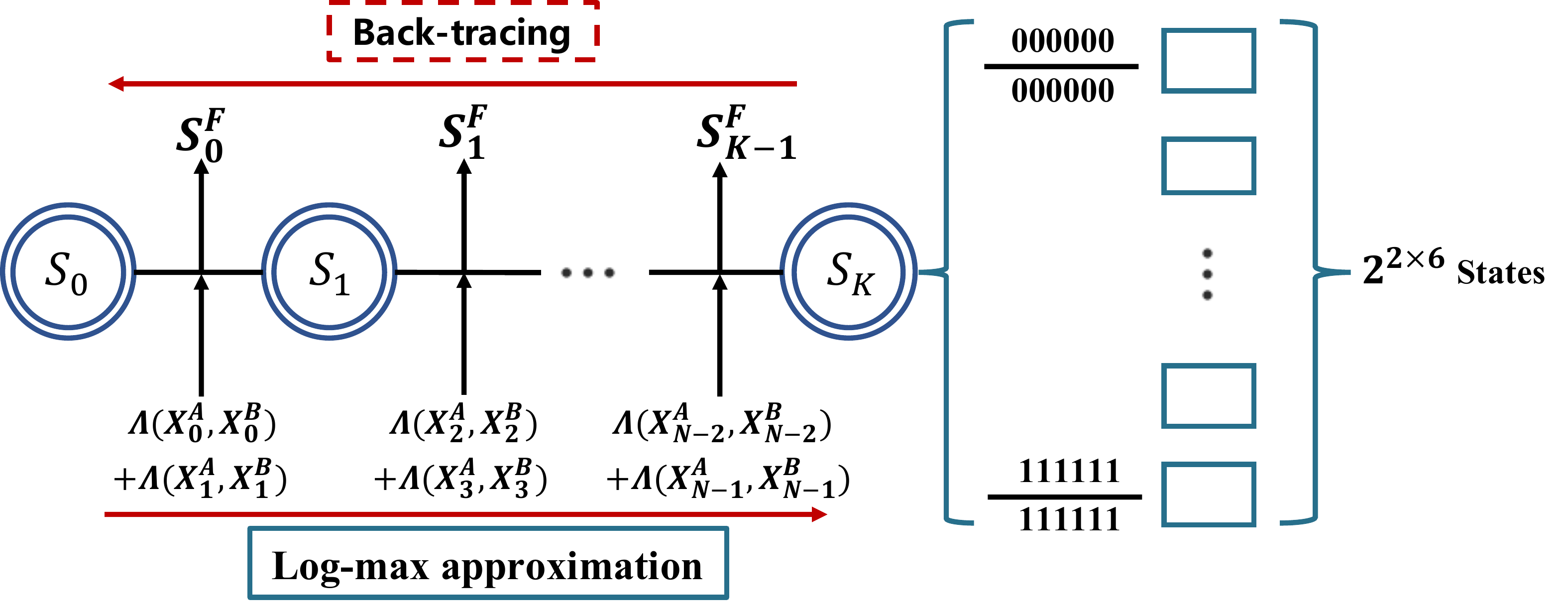}
	\caption{Joint trellis of the two-user convolutional decoder, assuming $K$ source bits, $N$ coded bits, 1/2 coding rate and constraint length $L$=7 for each user.}\label{fig:trellis}
\end{figure}

\subsection{Full-State Joint Decoder}\label{sec:conv:fsjd}

Full-state joint decoder (FSJD) is an approximated decoder for the ML decoder. In particular, Eq.~(\ref{eq1}) can be simplified using log-max approximation (i.e., $log(\sum_i exp(x_i)) \approx \max_i x_i$) to 

\begin{small}
\begin{equation} \label{eq3}
	\begin{split}
		\hat{S}^{F} &\approx \operatorname*{arg\,max}_{S^{F}} \max_{C^A, C^B: \Pi^{-1}(C^A)+\Pi^{-1}(C^B)=S^F} Pr(Y|C^A,C^B) \\
		&\approx \operatorname*{arg\,max}_{C^A, C^B: \Pi^{-1}(C^A)+\Pi^{-1}(C^B)=S^F} Pr(Y|C^A,C^B) \\
		&\approx \operatorname*{arg\,min}_{C^A, C^B: \Pi^{-1}(C^A)+\Pi^{-1}(C^B)=S^F}  \frac{\Lambda(X^A,X^B)}{2 \sigma^2} \\
		&\approx \operatorname*{arg\,min}_{C^A, C^B: \Pi^{-1}(C^A)+\Pi^{-1}(C^B)=S^F}  \Lambda(X^A,X^B) \\
            &\approx \operatorname*{arg\,min}_{C^A, C^B: \Pi^{-1}(C^A)+\Pi^{-1}(C^B)=S^F}  \sum_{n=1}^N \Lambda(X_n^A,X_n^B),
	\end{split}
\end{equation}
\end{small}
where
\begin{small}
\begin{equation} \label{eq4}
	\Lambda(X_n^A,X_n^B)=\Vert Y[n]-H^A[n]X^A[n]-H^B[n]X^B[n] \Vert^2.
\end{equation}
\end{small}

The computation of Eq.~(\ref{eq3}) includes two steps: 1) find the best pair of codewords $\hat{C}^A$ and $\hat{C}^B$ such that $(\hat{C}^A,\hat{C}^B) = \operatorname*{arg\,min}_{C^A,C^B} \Lambda(X^A,X^B)$; 2) map $\hat{C}^A$ to $\hat{S}^A$ and $\hat{C}^B$ to $\hat{S}^B$, and get $\hat{S}^F = \hat{S}^A + \hat{S}^B$. Step 1) is equivalent to finding the minimum-cost path on the joint trellis of user A's and user B's encoders, and can be implemented by using the Viterbi algorithm. In Eq.~(\ref{eq4}), for BPSK-modulated received signal $Y[n]$, we can construct four possible constellations with $H^A[n]$ and $H^B[n]$ ($X^A[n],X^B[n] \in \{+1,-1\}$), and calculate its Euclidean distance to the received point as the edge cost. The demodulation is called \emph{soft demodulation}.
Since the state space of the joint trellis is the combination of user A's and user B's state space, it is referred to as \emph{full-state} joint Viterbi decoder.

For single-user convolutional codes with constraint length $L$, the number of state in trellis is $2^{(L-1)}$. Each state branches out two edges to two states in the next stage, and each edge is associated with $1$ input bit and $r$ output bits for a convolution code with coding rate $r$. For the two-user joint trellis case, the number of states is $2^{2(L-1)}$, and each state branches out four edges to four states in the next stage. Moreover, each edge is associated with $2$ input bits (each user owns $1$ bit) and $2r$ output bits (each user owns $r$ bits). Fig.~\ref{fig:trellis} shows the joint trellis for two-user simultaneous transmission, where the joint state is represented by $2(L-1)$ bits with each user $L-1$ bits.

The Viterbi decoding algorithm is to find the minimum-cost path from the start state to the end state. In particular, for each decoding state, it performs the Add-Compare-Select (ACS) operation to compute and choose the minimum-cost path to the current state (i.e., path metric), and records the previous best state for back-tracing. The path metric value of the $k$-th state at the $i+1$-th decoding stage can be represented as:
\begin{equation}\label{PM}
	PM[k,i+1] =\underset{k'\in{set(k)}}{min}\left(PM[k',i]+BM[k'\rightarrow{k}]\right),
\end{equation}
where $set(k)$ is the set of four states that branch out to state $k$ at the last stage $i$, and $BM[k' \rightarrow k]$ is the branch metric from state $k'$ to $k$. Note that $BM[k' \rightarrow k]$ is the sum of corresponding bits' Euclidean distance in the constellation graph, while the distance for one bit is shown in Eq.~(\ref{eq4}).

After running the algorithm to the last decoding stage, we can determine the end state and identify the minimum-cost path from the start state to the end state. By tracking back the path, we can recover the joint source bits corresponding to the minimum-cost path and obtain the sum bits by summing the decoded source bits. The sum bits are then transformed back to floating-point values corresponding to each parameter, and average is performed to get the update model parameter value.
Fig.~\ref{fig:trellis} also shows the decoding procedure: by back-tracking, the decoder finds the minimal-cost path, and the found path corresponds to the optimal pair $(C^A, C^B)$; by mapping the pair to corresponding source bits $(S^A, S^B)$, the decoder gets $S^F$.

\subsection{Reduced-State Joint Decoder}\label{sec:conv:rsjd}
The computational complexity of FSJD is proportional to the number of states in each decoding stage, since each state is associated with one ACS operation. The aforementioned FSJD has a high computational complexity, since there are $2^{2(L-1)}$ states in each decoding stage. For the convolution codes defined in WiFi, the constraint length is 7, and there are 4096 states for two-user simultaneous transmissions. The state number increases exponentially as the user number $n$ increases (e.g., $2^{n(L-1)}$), leading to a high computation cost. For example, there are $2^{6\times3}=262144$ states for three-user simultaneous transmissions.

Reduced-state joint decoder (RSJD) is a simplified (and approximated) version of FSJD. In particular, it limits the number of states participating in the ACS operation in the next decoding stage. At most $R$ states with minimum path metric are selected to advance to the next decoding stage, and other states are ignored. Therefore, we call it \emph{reduced-state} joint Viterbi decoder.

In our joint trellis, $R$ states branch out to at most $2^nR$ states in the next decoding stage, but RSJD still selects at most $R$ states for the following decoding stages. We use the quick-sort algorithm to select these $R$ states in our implementation. In this way, RSJD reduces the computational complexity, although it may not be the optimal in terms of the minimum-cost path, since some paths with smaller cost may be ignored. 

In Section \ref{sec:eva:sum-ber}, we study the performance loss under two-user simultaneous transmissions. Results show that the performance of RSJD approaches that of FSJD for most SNRs, especially for high SNR regimes. The performance loss may be larger for more users, since the total number of states increases exponentially. However, such reduced-state approximation is essential for real-time implementation.  Therefore, we adopt RSJD and adapt the number of reserved states according to the number of simultaneous transmissions in our implementation.

\rev{
\subsection{Complexity Analysis}
We have presented an optimal decoder and two reduced-complexity decoders. In the following, we give a complexity analysis of these decoders, and show how much complexity can be reduced.

First, the optimal decoder can also be viewed as computations in the joint trellis. Different from the Viterbi algorithm that only keeps one path in each state, the optimal decoder keeps all possible paths in each state. Therefore, we can focus on the number of performed ACS operations, and ignore the computation of branch metrics that are common for all three decoders.

For the general $M$-user digital AirComp system, the optimal decoder needs to consider $2^{MK}$ codeword pairs and $(M+1)^K$ SUM codewords. 
We need to exhaustively compute the probabilities for all codeword pairs, and sum the probabilities of corresponding pairs to compute the probability of a SUM codeword. Therefore, the optimal decoder needs $2^{MK}$ add operations, and 1 compare-select operation. Instead, FSJD needs $2^{M(L-1)} \cdot K$ ACS operations, and RSJD needs $R \cdot K$ ACS operations. If we ignore the complexity of the compare-selection operation, the computation complexity of these decoders are $O(2^{MK})$, $O(2^{M(L-1)} \cdot K)$, and $O(R \cdot K)$, respectively. Given the packet length $K$ is much larger than the constraint length $L$, FSJD reduces the complexity from exponential (i.e., $2^K$) to linear (i.e., $K$). RSJD further reduces the complexity from $2^{M(L-1)}$ to $R$.
}

\section{Design for LDPC Codes} \label{sec:ldpc} 
LDPC codes are linear block codes with a sparse generator matrix. Compared with convolutional codes, they have a better performance in approaching the Shannon capacity. 
For single-user communication systems, codeword-optimal (ML) decoding for LDPC codes is NP-hard, and the iterative belief propagation (BP) algorithm that achieves the bit-optimal performance is often adopted. 
Therefore, we also focus on the BP-based \emph{SUM bit-optimal decoder} to decode the arithmetic sum. The difficulty lies in defining the passing message and the update equations. This section present the decoding algorithm in details.

\subsection{SUM Bit-optimal Decoder}

The SUM bit-optimal decoder is defined as follows.
The $i$-th SUM source bits $\hat{S}^F_i$, $i=1,\cdots,K$, is given by 
\begin{equation} \label{eqn:bit-optimal}
    \hat{S}^F_i = \operatorname*{arg\,max}_{S^{F}_i} \sum_{S^A_i, S^B_i: S^A_i+S^B_i=S^F} Pr(S_i^A,S_i^B|Y,\mathcal{C}),
\end{equation}
where $Pr(S_i^A,S_i^B|Y,\mathcal{C})$ denotes the \emph{a posteriori} probability of $(S_i^A,S_i^B)$ given the received signal $Y$ and the codebook $\mathcal{C}$.
Eq.~(\ref{eqn:bit-optimal}) is to compute the probability of $Pr(S^A_i, S^B _i |Y, \mathcal{C})$, map $(S^A_i, S^B _i)$ to $S^F_i$ according to the functional mapping method
$$f_{bit}: \{0, 1\} \times \{0, 1\} \rightarrow \{0,1,2\},$$
and choose the $S^F_i$  value with the maximal probability. It is a (4/3)-to-1 mapping, and the mapping complexity is significantly lower than the codeword-optimal decoding.

Then, the key is to compute the marginal probability distributions $Pr(S^A_i, S^B_i |Y, \mathcal{C})$. That is, we compute the probability of $(S^A_i, S^B_i)$ with the assistance of coding structure in codebook $\mathcal{C}$. Fortunately, it can be computed using the classical BP framework, which is a general framework for generating inference-making algorithms for graphical models. It can find the ML SUM source bit without incurring exponential growth in complexity. %

\subsection{Bit-optimal LDPC Decoder} \label{sec:ldpc:decoding}
LDPC codes are linear block codes with a sparse generator matrix. Compared with convolutional codes, they have a better performance in approaching the Shannon capacity. 
For single-user communication systems, codeword-optimal (ML) decoding for LDPC codes is NP-hard, and the iterative belief propagation (BP) algorithm that achieves the bit-optimal performance is often adopted. 
Therefore, we focus on the BP-based \emph{SUM bit-optimal decoder} to decode the arithmetic sum \rev{for LDPC codes\footnote{\rev{There also exists a SUM bit-optimal decoder for convolutional codes. However, its performance does not make a big difference compared with the approximated codeword-optimal decoders in Section \ref{sec:conv} despite higher complexity. Therefore, we do not show the design here, and put the details in Appendix C.}}}. The difficulty lies in defining the passing message and the update equations. This section present the decoding algorithm in details.

Our multi-user LDPC SUM decoder reuses the Tanner graph structure for single-user LDPC, but redesigns the passing message and the update rules for decoding. For ease of illustration, we use the two-user LDPC SUM decoder for example. The extension to more users is straightforward.

\textbf{Message Vector:}  The SUM decoder uses a message vector that contains the probability of multi-users' bit vector instead of a log likelihood ratio message for single-user. For example, for two-user LDPC, we define the information passed between nodes as a vector $(p_{00}, p_{01}, p_{10}, p_{11})$, in which 
\begin{equation}
    p_{ij}=Pr(X^A[n]=2i-1,X^B[n]=2j-1|Y[n])    
\end{equation}
is the probability that the user A's $n$-th coded bit is $i$ and the user B's $n$-th coded bit is $j$ given the $n$-th received bit signal assuming  the BPSK bit-to-signal mapping of $1 \rightarrow 1$ and $0 \rightarrow -1$. The message vector will be used to update the information in both variable nodes and check nodes. 

\begin{figure}
	\centering
	\subfloat[Tanner graph.]{  
		\includegraphics[width=3.4 in]{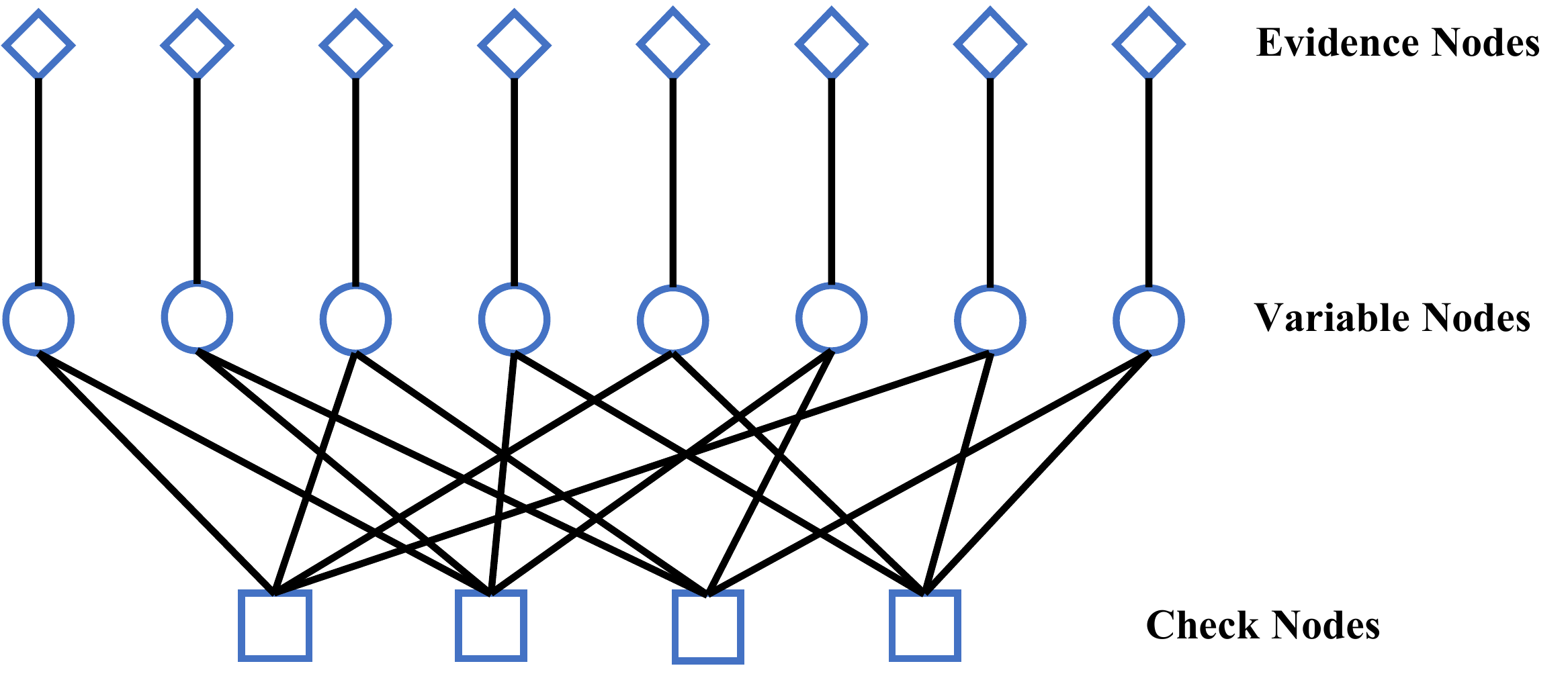}
        \label{fig:ldpc:tanner}
    }
	\quad
	\subfloat[Update on a check node.]{
		\includegraphics[width=1.5 in]{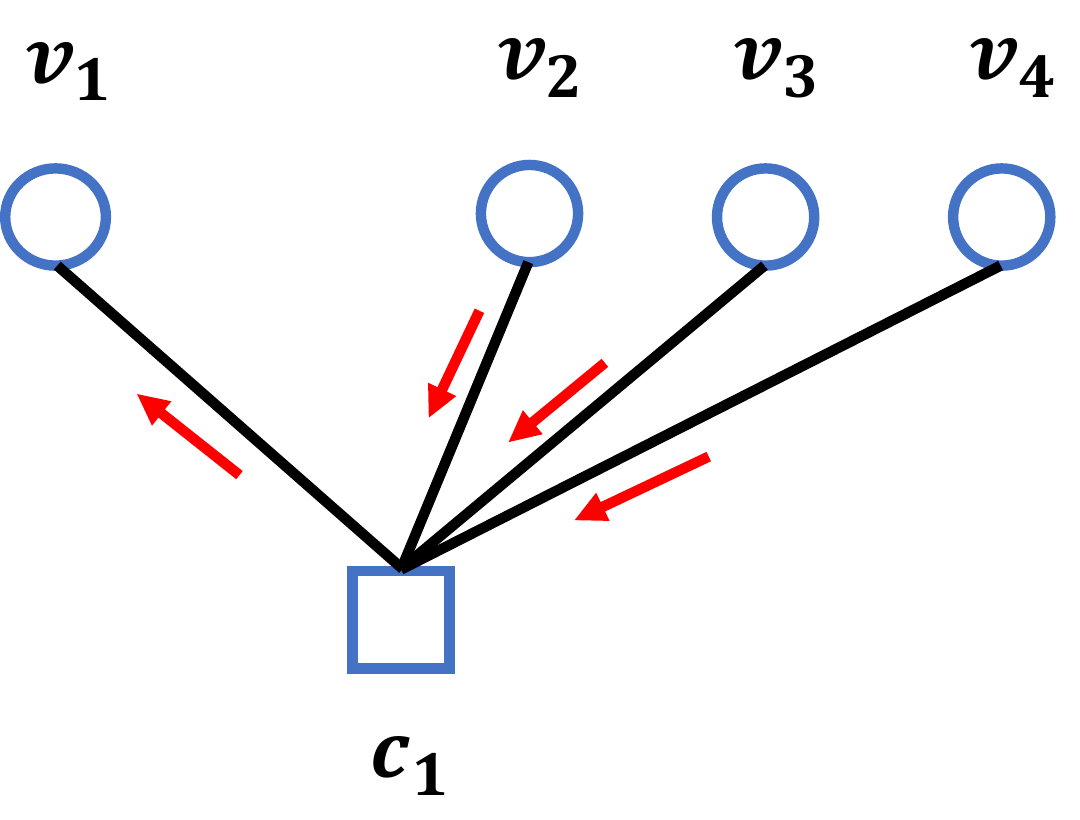}
        \label{fig:ldpc:chk}
    }
    \hspace{0.1 in}
	\subfloat[Update on a variable node.]{
		\includegraphics[width=1.5 in]{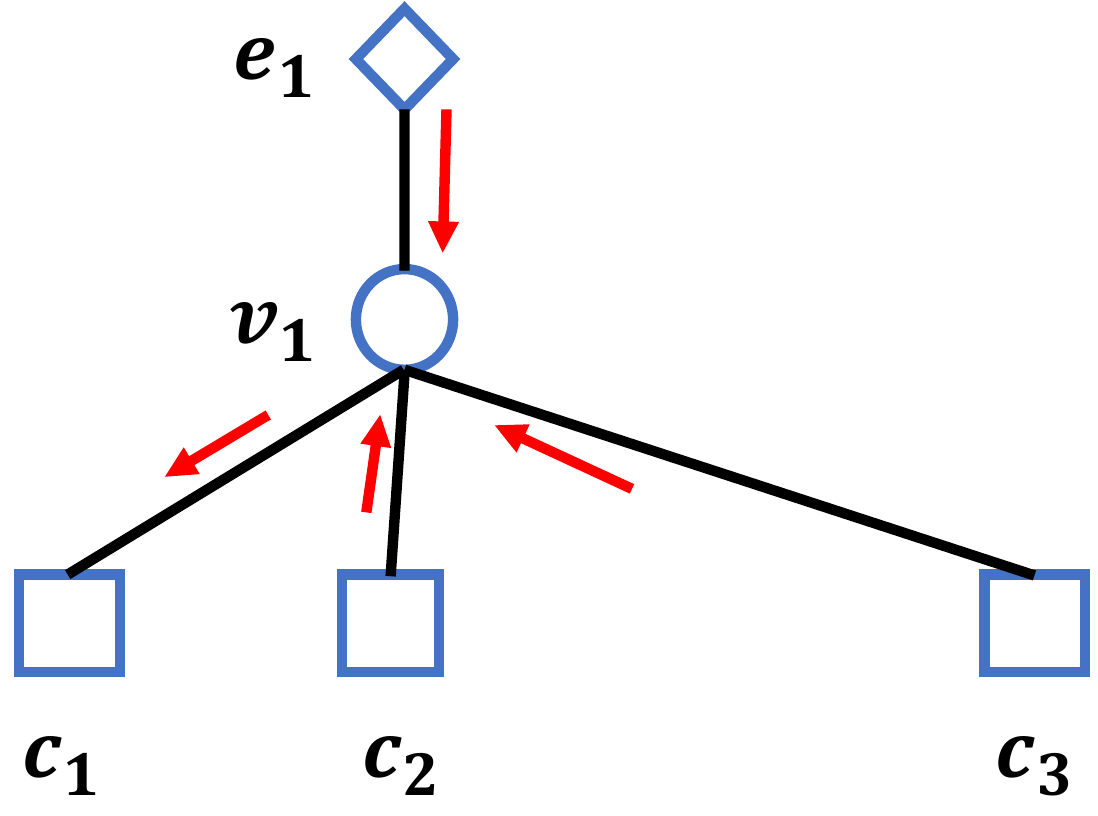}
        \label{fig:ldpc:var}
    }
	\caption{LDPC decoder for multi-user simultaneous transmissions.}
	\label{fig:ldpc}
\end{figure}

\textbf{Tanner Graph:} Fig.~\ref{fig:ldpc}(a) shows the Tanner graph for the multi-user simultaneous transmission system. Given the above message vector definition, we can reuse the Tanner graph for single-user LDPC. A variable node correspond to an output bit vector, and the check nodes correspond to their redundancy relationship that their finite-field sum is zero. The evidence nodes correspond to the received signal associated with variable nodes.

\textbf{Message Initialization:} %
The information from an evidence node to a variable node is the initial messages. Note that the message remains the same in every iteration of the decoding process.
The information from the evidence node $n$ is computed from the received signal $Y[n]$ as
\begin{equation}
    \begin{aligned}
        P&=(p_{00},p_{01},p_{10},p_{11})\\
        &=\frac{1}{\beta_1} (Pr(X^A[n]=-1,X^B[n]=-1|Y[n]),\\
        &\quad\enspace  Pr(X^A[n]=-1,X^B[n]=1|Y[n]), \\
        &\quad\enspace  Pr(X^A[n]=1,X^B[n]=-1|Y[n]), \\
        &\quad\enspace  Pr(X^A[n]=1,X^B[n]=1|Y[n]))\\
        &=\frac{1}{\beta_1}(exp(\frac{-D(-1,-1)}{2\sigma^2}),exp(\frac{-D(-1,1)}{2\sigma^2}), \\
        &\quad\enspace exp(\frac{-D(1,-1)}{2\sigma^2}),exp(\frac{-D(1,1)}{2\sigma^2})),
    \end{aligned} \label{eqn:message:init}
\end{equation}
where $\beta_1$ is the sum of four probability for normalization and $D(X^A[n],X^B[n])$ is given by
\begin{equation}
    D(X^A[n],X^B[n])=\frac{\Vert Y[n]-H^A[n]X^A[n]-H^B[n]X^B[n] \Vert^2}{\alpha}. \nonumber
\end{equation}
Here $\alpha$ represents the Euclidean distance normalization factor, since the power of superimposed signal in different subcarriers may be diverse due to heterogeneous phase offsets. 
We set $\alpha$ as the minimal distance over all subcarriers in one packet. 

\textbf{Update Equations on Check Nodes:} We use function $CHK(\cdot)$ to denote the update equation that computes the output message $\mathcal{M}(\cdot)$ going out of a check node to a variable node. $CHK(\cdot)$ supports multiple inputs, and is computed sequentially, i.e., $$CHK(U,V,\cdots)=CHK(CHK(U,V),\cdots).$$
Fig.~\ref{fig:ldpc}(b) shows an example of the updating process on a check node $c_1$.
It is to compute $\mathcal{M}(c_1 \rightarrow v_1)$ defined as $$\mathcal{M}(c_1 \rightarrow v_1)=CHK(\mathcal{M}(v_2 \rightarrow c_1), \mathcal{M}(v_3 \rightarrow c_1), \mathcal{M}(v_4 \rightarrow c_1)).$$ 
To ease notation, we rewrite $\mathcal{M}(c_1 \rightarrow v_1)$ as $CHK(v_2 \rightarrow c_1, v_3 \rightarrow c_1, v_4 \rightarrow c_1)$ or $CHK(\{v_2,v_3,v_4\} \rightarrow c_1)$.

Since $CHK(\cdot)$ is computed sequentially, we focus on the minimal computation unit $CHK(I, J)$, where the message vectors of the two input information edges are $I=(i_{00},i_{01},i_{10},i_{11})$ and $J=(j_{00},j_{01},j_{10},j_{11})$. 
The probability that the composed temporary node $v$ is 00 is obtained as
\begin{equation}
    \begin{aligned}
        Pr&(v=00|I,J)\\
        &= Pr(i=00,j=00|I,J)+Pr(i=01,j=01|I,J)+\\
        &\quad\enspace Pr(i=10,j=10|I,J)+Pr(i=11,j=11|I,J)\\
        &=i_{00}j_{00}+i_{01}j_{01}+i_{10}j_{10}+i_{11}j_{11}.
    \end{aligned} \nonumber
\end{equation}
Similarly, we can obtain $Pr(v=01|I,J),Pr(v=10|I,J)$ and $Pr(v=11|I,J)$. Then the output message vector is given by
\begin{equation}
    \begin{aligned}
        CHK(I,J)= (&i_{00}j_{00}+i_{01}j_{01}+i_{10}j_{10}+i_{11}j_{11},\\
        &i_{00}j_{01}+i_{01}j_{00}+i_{10}j_{11}+i_{11}j_{10},\\
        &i_{00}j_{10}+i_{10}j_{00}+i_{01}j_{11}+i_{11}j_{01},\\
        &i_{00}j_{11}+i_{11}j_{00}+i_{01}j_{10}+i_{10}j_{01}).
    \end{aligned} \label{eqn:check}
\end{equation}
Then we can apply $CHK(\cdot)$ to include more input variable nodes and obtain the final output message. %

In this way, we use $CHK(\cdot)$ to obtain the output message from the check node to one variable node. Similarly, we can use $CHK(\cdot)$ to compute output messages from the check node to other variable nodes. %

\begin{algorithm}[t!]
	\label{algo:bp}
	\LinesNumbered
    \KwIn{Number of users: $M$; Received baseband signal in frequency domain: $Y$; Check nodes list: $C$; Variable nodes list: $V$%
    .}
    \KwOut{Arithmetic sum of source bits: $\hat{S}^F$.}
	Calculate messages out of evidence nodes $\mathcal{M}(e_i \rightarrow v_i)$ according to Eq.~(\ref{eqn:message:init})\;
	Initialize messages from variable nodes to associated check nodes: $\mathcal{M}(v_i \rightarrow c_j) = \mathcal{M}(e_i \rightarrow v_i)$\;
    \While{iter $<$ MAX-ITER}{	
        \For{check nodes $c_i$ in $C$}{
            $V_i \leftarrow$ variable nodes linked with $c_i$\;
            \For{variable node $v_j \in V_i$}{
                $\mathcal{M}(c_i \rightarrow v_j) = CHK(\mathcal{M}(V_i \setminus v_j \rightarrow c_i))$\;
            }
        }
        \For{variable nodes $v_i$ in $V$}{
            $C_i \leftarrow$ check nodes linked with $v_i$\;
            \For{check node $c_j \in C_i$}{
                $\mathcal{M}(v_i \rightarrow c_j) = VAR(\mathcal{M}(e_i  \rightarrow v_i), \mathcal{M}(C_i \setminus c_j \rightarrow v_i))$\;
            }
        }
    }
    \For{variable nodes $v_i$ in $V$}{
        $C_i \leftarrow$ check nodes linked with $v_i$\;
        $P = p_{s_1 \cdots s_M} = VAR(\mathcal{M}(e_i \rightarrow v_i), \mathcal{M}(C_i \rightarrow v_i))$\; 
        Obtain the sum bit $s$: $\operatorname*{arg\,max}_{s=s_1+\cdots+s_M} \sum p_{s_1 \cdots s_M};$
    }
    Extract the SUM source bits $\hat{S}^F$ in selected variable nodes (for systematic codes).
    
\caption{BP algorithm for LDPC Jt-CDA.}
\end{algorithm}

\textbf{Update Equations on Variable Nodes:} We use function $VAR(\cdot)$ to denote the update equation that computes output message $\mathcal{M}$ going out of a variable node to a check node. Note that the output message is used as input to $CHK(\cdot)$. For multiple inputs, the computation of $VAR(\cdot)$ is also sequential. Fig.~\ref{fig:ldpc}(c) shows an example of the updating process on a variable node $v_1$. It is to compute $\mathcal{M}(v_1 \rightarrow c_1) = VAR(\mathcal{M}(e_1 \rightarrow v_1), \mathcal{M}(c_2 \rightarrow v_1), \mathcal{M}(c_3 \rightarrow v_1))$.

We focus on the minimal computation unit $VAR(I,J)$, where the message vectors of the two input information edges are $I=(i_{00},i_{01},i_{10},i_{11})$ and $J=(j_{00},j_{01},j_{10},j_{11})$. The probability that the composed temporary node $v$ is 00 is obtained as
\begin{equation}
    \begin{aligned}
        Pr(v=00&|I,J)=\frac{Pr(I,J|v=00)Pr(v=00)}{Pr(I,J)}\\
        &=\frac{Pr(I|J,v=00)Pr(J|v=00)Pr(v=00)}{Pr(I,J)}\\
        &=\frac{Pr(I|v=00)Pr(J|v=00)Pr(v=00)}{Pr(I,J)}\\
        &=\frac{Pr(v=00|I)Pr(v=00|J)Pr(I)Pr(J)}{Pr(I,J)Pr(v=00)}\\
        &=\alpha_2 i_{00}j_{00},
    \end{aligned} \label{eqn:var}
\end{equation}
where $\alpha_2=4\frac{Pr(I)Pr(J)}{Pr(I,J)}$ and the two input messages are assumed to be independent given the value of the variable node, i.e., $Pr(I|J,v)=Pr(I|v)$.

In a similar way, we can obtain $Pr(v=01|I,J)$, $Pr(v=10|I,J)$ and $Pr(v=11|I,J)$ as well. Thus, the output message at the variable node is
\begin{equation}
    \begin{aligned}
        VAR(I,J)&=\frac{1}{\beta_2}(i_{00}j_{00},i_{01}j_{01},i_{10}j_{10},i_{11}j_{11}),
    \end{aligned}
\end{equation}
where $\beta_2$ is the sum of four probability for normalization. Note that $\beta_2$ does not include $\alpha_2$, since $\alpha_2$ is a common factor that can be ignored during normalization.

In this way, we use $VAR(\cdot)$ to obtain the output message from the variable node to one check node. Similarly, we can use $VAR(\cdot)$ to compute output messages from the variable node to other check nodes.

\textbf{Overall Algorithm}: Algorithm~\ref{algo:bp} shows the BP algorithm for Jt-CDA in overall. Initially, we calculate the message vectors out of evidence nodes by received baseband signal $Y$. The message also initializes the message vectors from variables nodes to their associated check nodes. Then, following the update equations of Eq.(\ref{eqn:check}) and Eq.(\ref{eqn:var}), we can iterate the message passing process. After fixed rounds of iteration, we can calculate a probability vector in each variable node that represents the joint probability of each user's bit. Then, we can use the (4/3)-to-1 mapping to sum the probability for the SUM bit and determine the best SUM bit. For systematic LDPC codes, we can directly extract SUM bits in corresponding positions to get $\hat{S}^F$.

\section{Evaluation} 
\label{sec:results}

\begin{figure}[t!]
	\centering
	\subfloat[]{
		\includegraphics[width=1.12in]{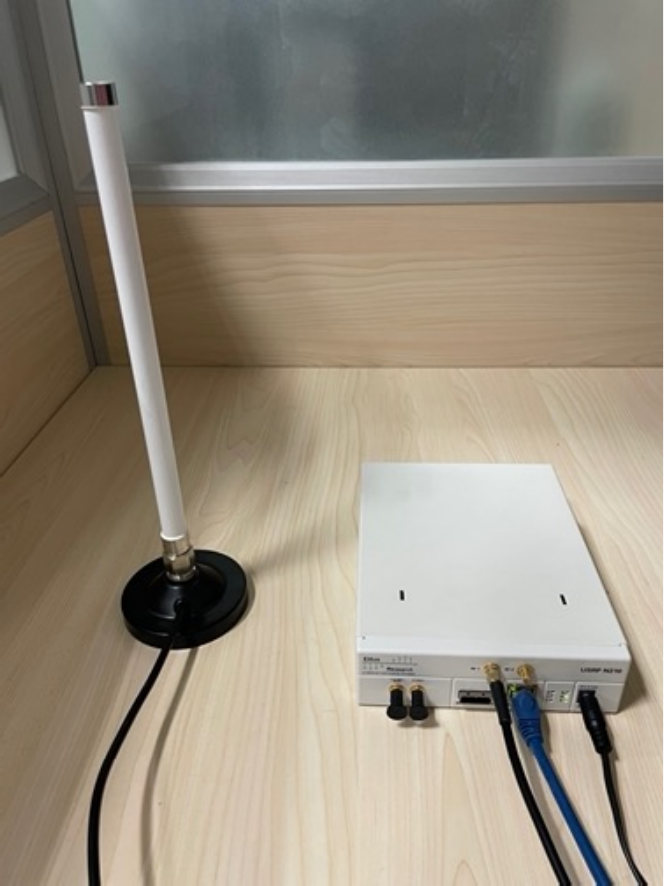} %
		\label{fig:params_server}}
        \hspace{0.1in}
	\subfloat[]{
		\includegraphics[width=2in]{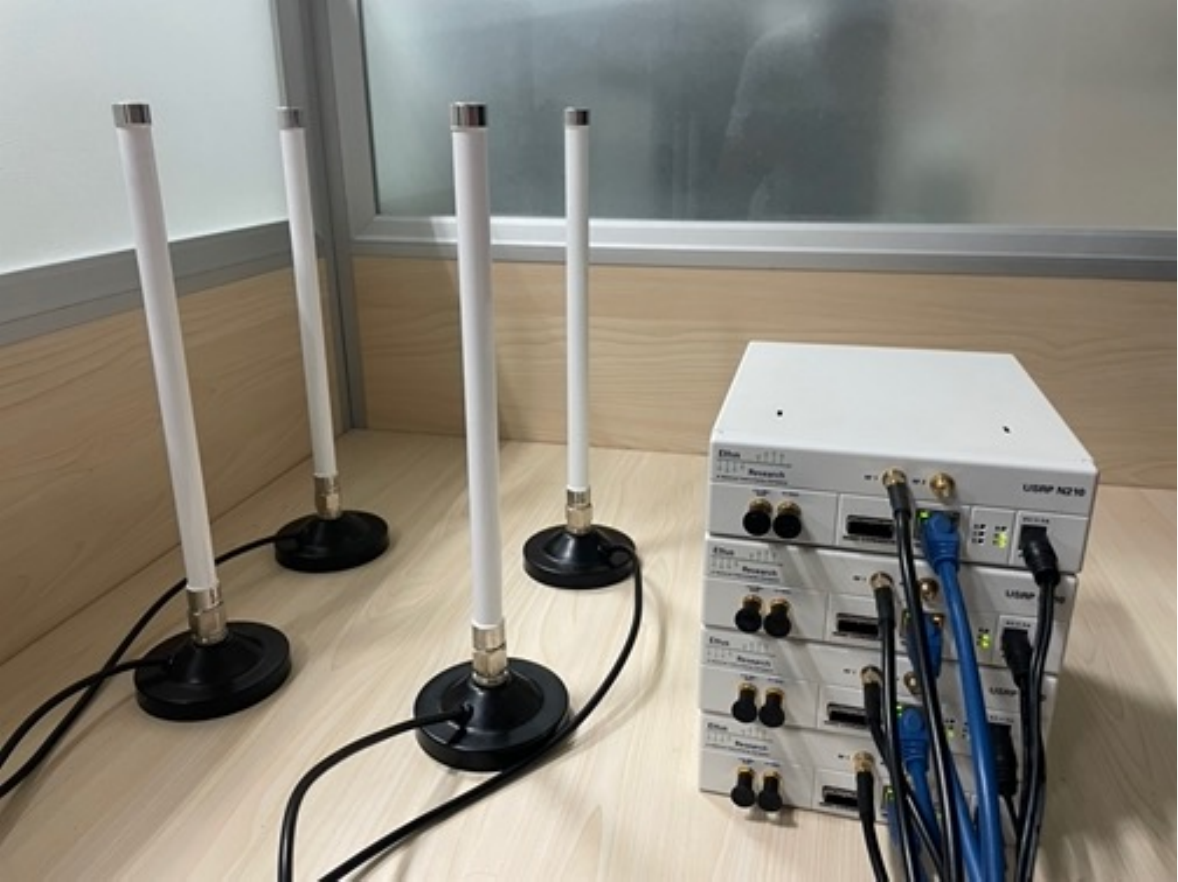} %
		\label{fig:edge_device}}
	\caption{A digital AirComp testbed on the USRP platform: (a) an edge parameter server; (b) four edge devices.}
	\label{fig:testbed}
\end{figure}

\begin{figure*}[t!]
	\centering
	\subfloat[]{
		\includegraphics[width=2.25in]{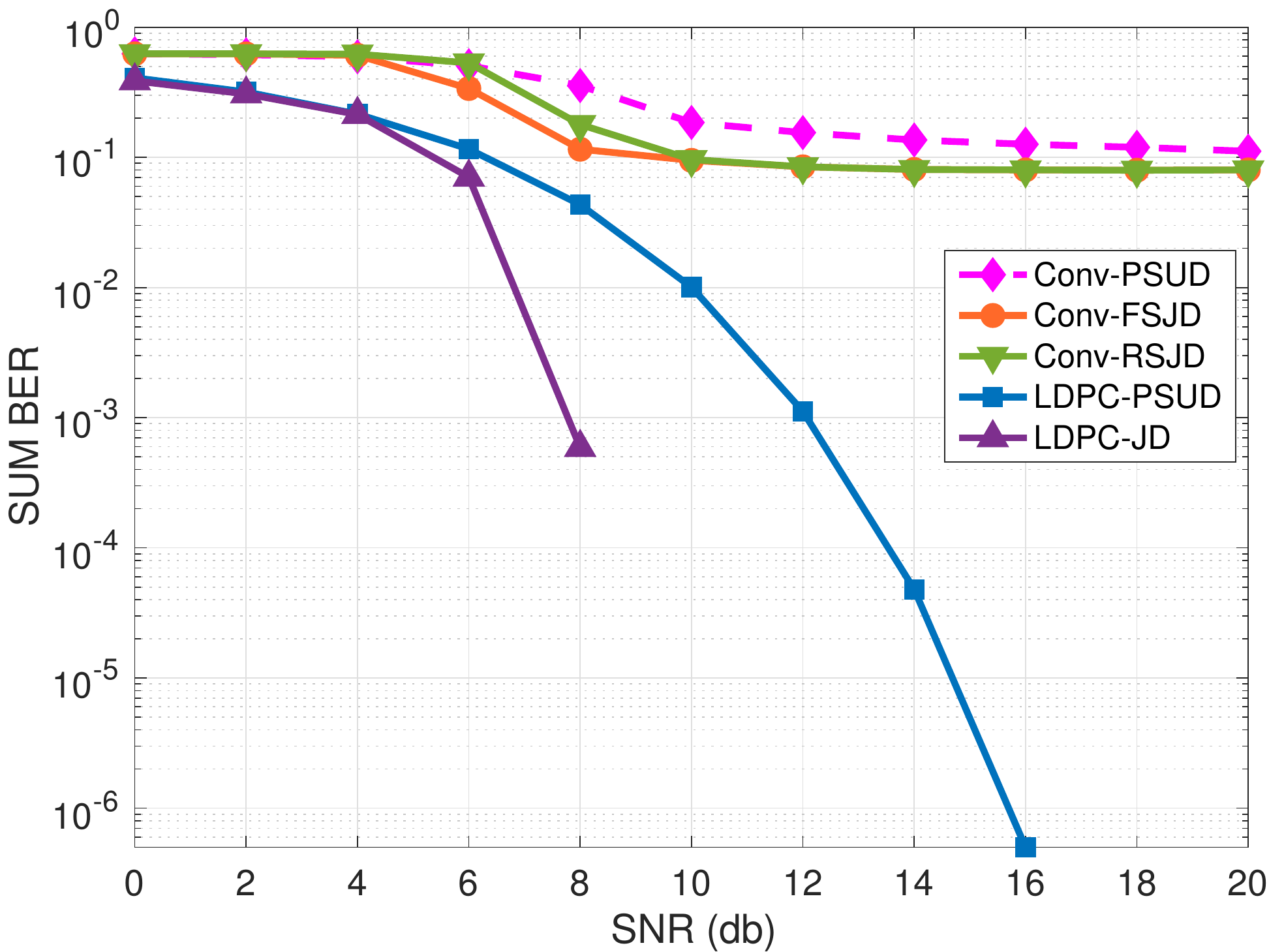} %
		\label{fig:phase0}}
	\subfloat[]{
		\includegraphics[width=2.25in]{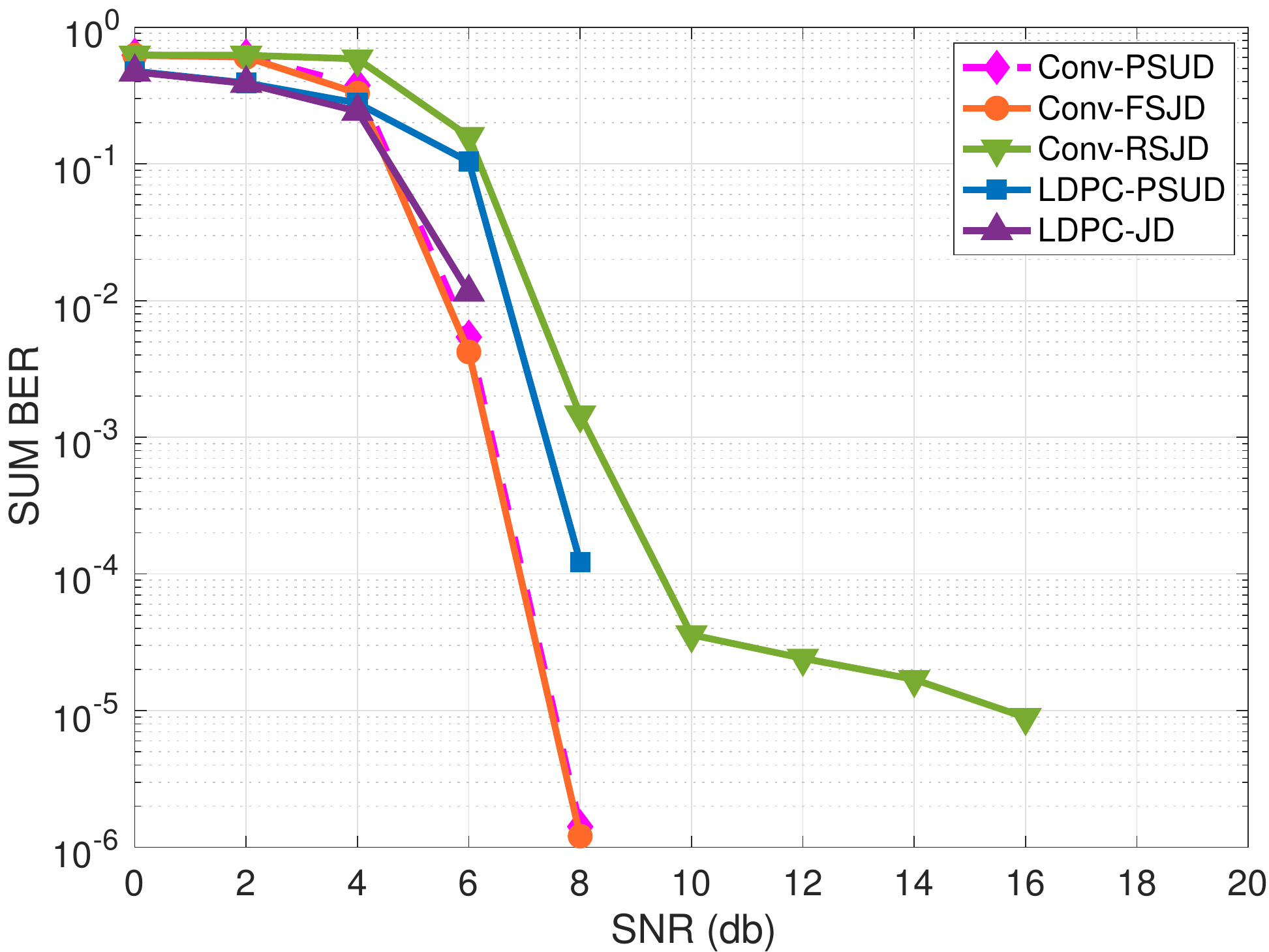} %
		\label{fig:phase90}}
	\subfloat[]{
		\includegraphics[width=2.25in]{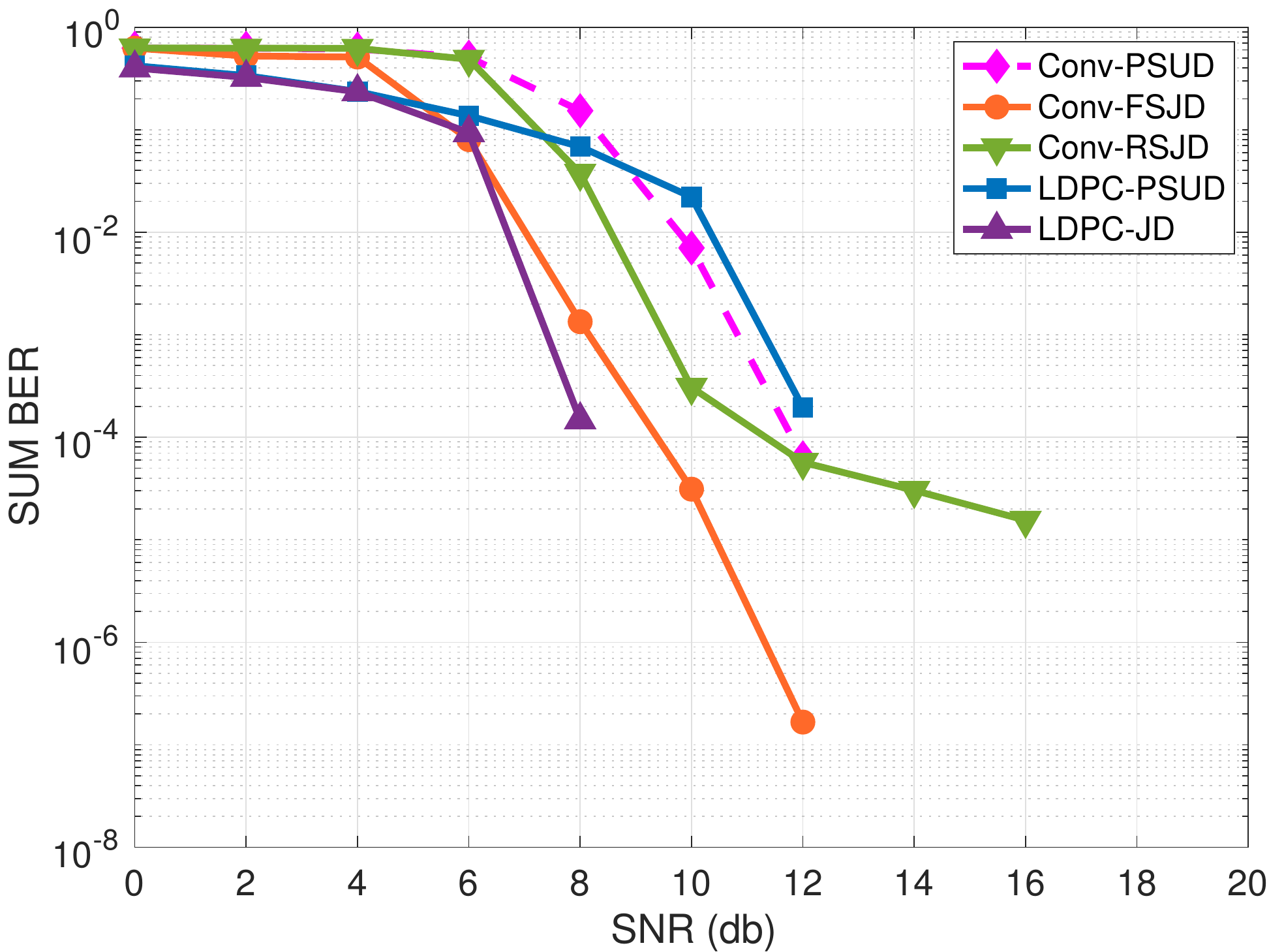} %
		\label{fig:phase45}}
	\caption{SUM BER versus SNR under various phase settings in AWGN channel for two-user simultaneous transmissions: (a) perfectly aligned channel phase; (b) relative channel phase offset of $\pi$/2 radian; (c) relative channel phase offset of $\pi$/4 radian.}
	\label{fig:BER_SNR}
	\vspace{-2ex}
\end{figure*}

To test our digital AirComp systems in a practical environment, we use the USRP software-defined radio as the experiment platform. We implement a testbed using five USRP N210s with UHD driver 3.8.2 as shown in Fig.~\ref{fig:testbed}: four USRPs serve as edge devices, and one USRP serves as the parameter server. USRPs are equipped with the UBX daughterboards. All experiments are performed at 5.85GHz with a bandwidth 10MHz.

To support simultaneous transmissions, we use the implemented MAC protocol in \cite{Lu2013}: a beacon triggers simultaneous transmissions; edge devices detect the beacon with precise hardware timestamp on USRP (on the order of \emph{ns}), and add sufficient large time that compensates the signal transfer time and the baseband processing time for transmission. Meanwhile, edge devices measure CFOs according to the beacon frame in the downlink, and perform phase precoding to reduce the CFO in the uplink. We measure the synchronization accuracy under the current hardware and protocol: TO is less than $2$ samples, and CFO is within $[-2,+2]$ kHz.

We implement the transmitter and the receiver shown in Fig.~\ref{fig:arch} using GNU Radio 3.7.11. GNU Radio is a real-time signal processing framework written in C++ and Python. The core signal processing functionalities are all implemented in C++ for real-time operation. Each block in Fig.~\ref{fig:arch} is implemented as a GNU Radio block. Signal processing blocks are run in parallel using multiple threads, and are scheduled by the operating system using a default policy specified by GNU Radio.
Each OFDM frame includes 500 data symbols. We adopt BPSK and 1/2 coding rate, and thus each OFDM frame has around 1500 Bytes, close to the maximal length of an Ethernet frame.
For convolutional-coded AirComp, we adopt the code defined in WiFi standards with polynomials $(133_8, 171_8)$. For LDPC-coded AirComp, we adopt the block-structured code defined in WiFi standards with each block 1296 bits.%

\begin{figure*}[t!]
	\centering
	\subfloat[]{
		\includegraphics[width=2.25in]{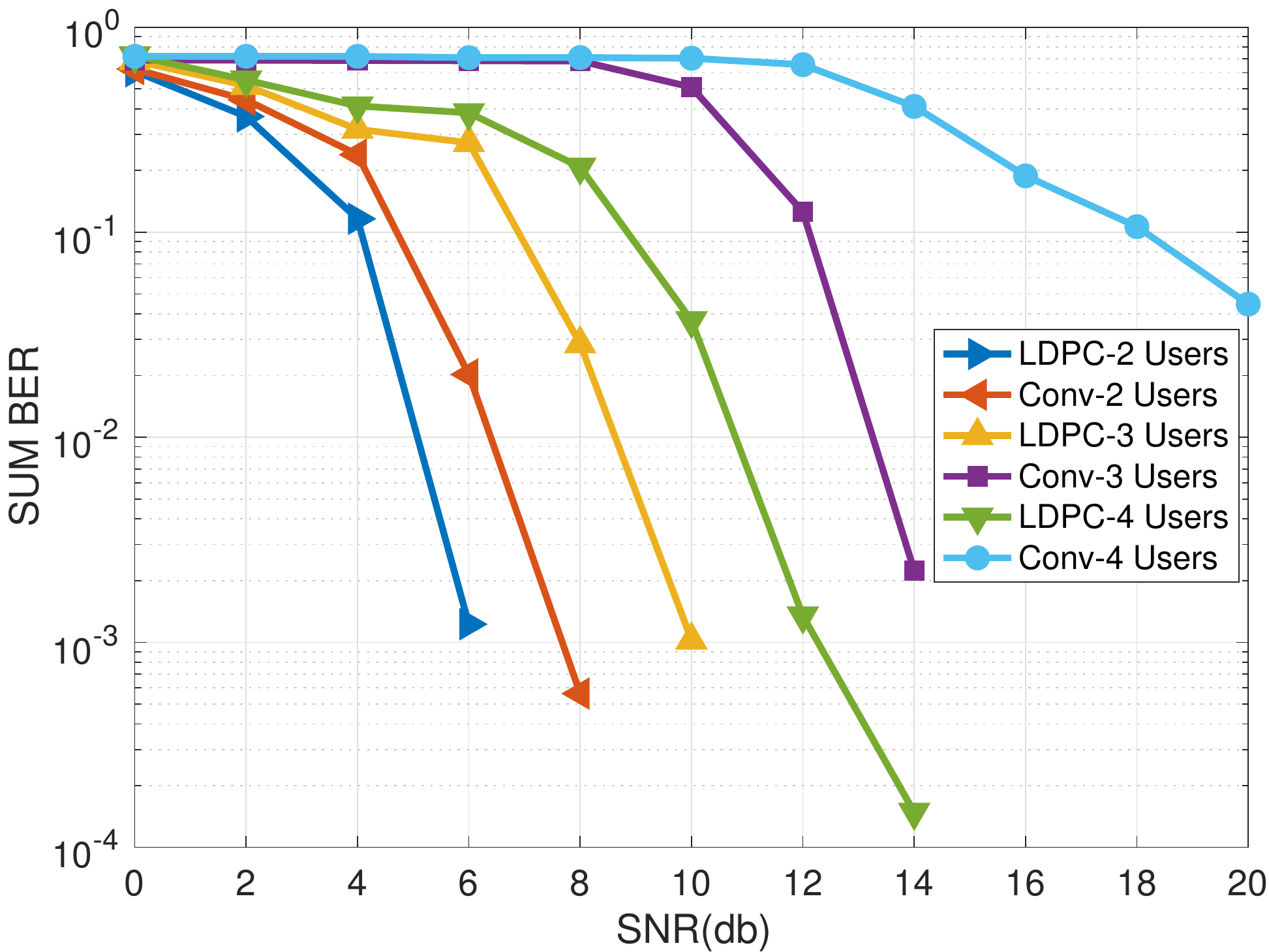} %
		\label{fig:usrp_imp}}
	\subfloat[]{
		\includegraphics[width=2.25in]{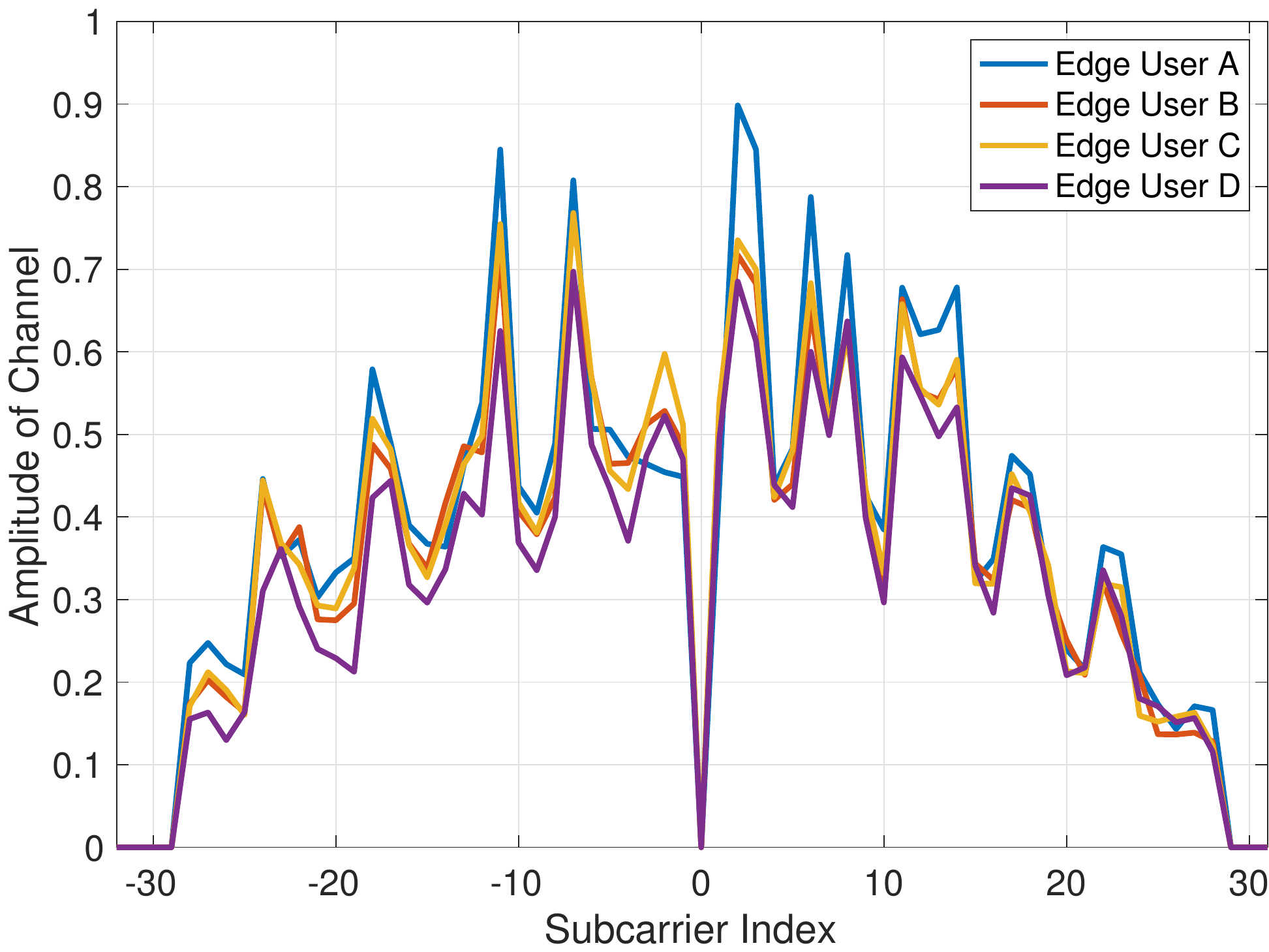} %
		\label{fig:channel_amp}}
	\subfloat[]{
		\includegraphics[width=2.25in]{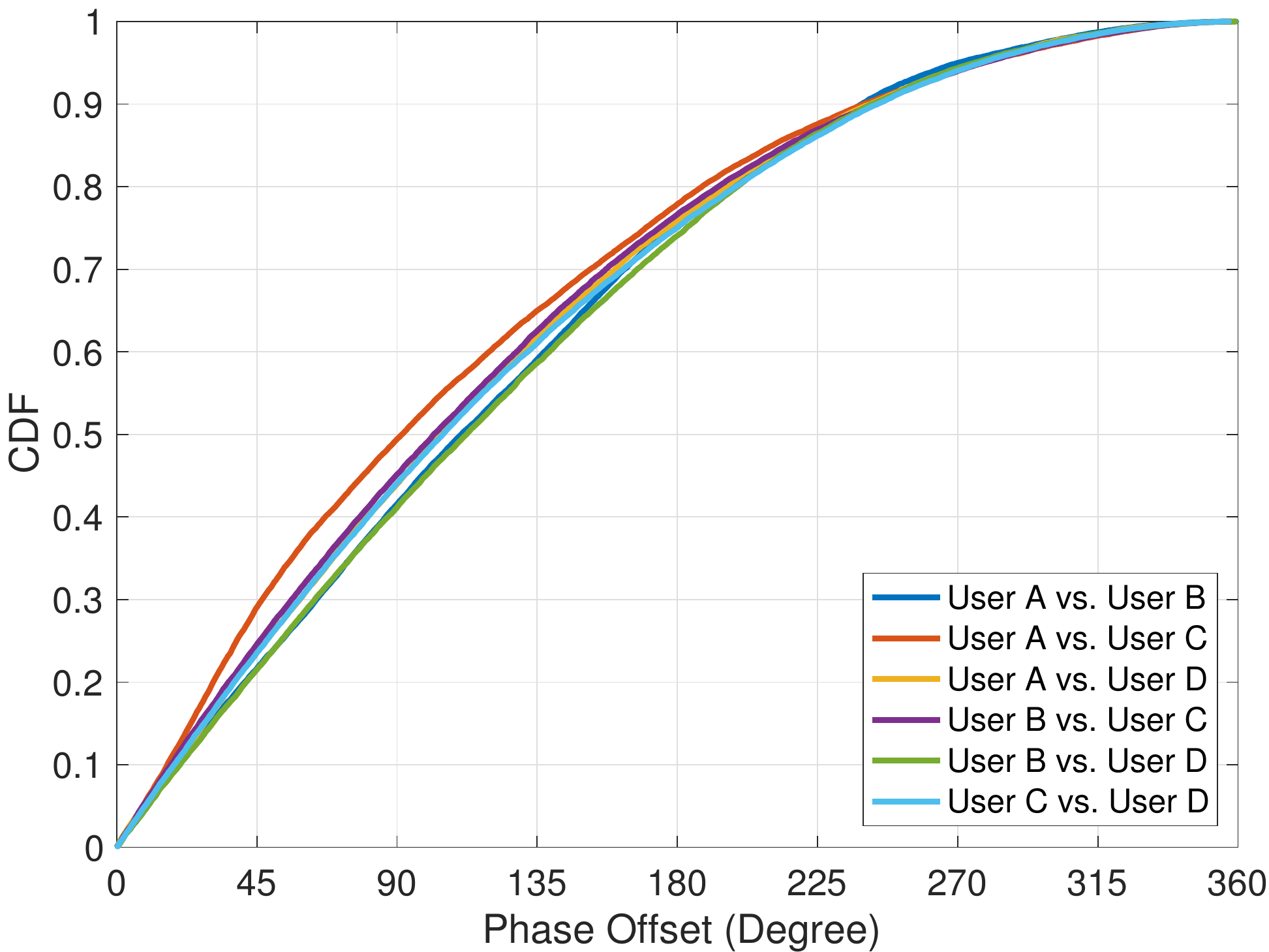} %
		\label{fig:phase_offset}}
	\caption{(a) SUM SER versus SNR under experimental channel on USRP; (b) channel amplitude of transmission users in an experiment; (c) cumulative distribution function of relative phase offset between transmission users in an experiment.}
	\label{fig:BER_SNR_REAL}
\end{figure*}

\subsection{SUM BER Performance} \label{sec:eva:sum-ber}
\textbf{Setup.}
We use the sum bit error rate (SUM BER) as a metric to evaluate the decoding performance of the proposed Jt-CDA decoders. %
For two devices, there are three possible SUM bit outcomes $\{0,1,2\}$, and any mismatch is treated as an error. We compare their SUM BER performance for different SNRs (less than 20 dB). For each SNR, we run 4000 times, and present the average result. SNR is defined as the reception power of the superimposed signal versus noise. %

We consider the following proposed joint decoders: \emph{Conv-FSJD}, \emph{Conv-RSJD}, and \emph{LDPC-JD}. 
Conv-FSJD and Conv-RSJD belong to \emph{Conv-JD}.
For Conv-RSJD, we consider $R$=256, 1024, 4096 for two-user, three-user and four-user simultaneous transmissions, respectively.
Meanwhile, we also consider two parallel separate decoders (PSUD), \emph{Conv-PSUD} and \emph{LDPC-PSUD} for comparison. Different from joint decoders, PSUDs perform symbol-level probability marginalization to separate and decode each user's data, and thus also have low implementation complexity. Appendix D shows their formal definitions. %

To study the impact of relative phase offset on all Jt-CDA decoders, we consider the following ideal channel conditions with two-user simultaneous transmissions, and generate superimposed signals using simulations:
\begin{enumerate}
    \item AWGN channel with unit amplitude and no relative channel phase offset. %
    \item AWGN channel with unit amplitude and relative channel phase offset of $\pi/2$ radian. %
    \item AWGN channel with unit amplitude and relative channel phase offset of $\pi/4$ radian. %
\end{enumerate}
We also consider an experimental realistic channel using USRPs. We use the aforementioned MAC protocol to trigger simultaneous transmissions, and log the over-the-air superimposed baseband signals. Then we run various decoders on the same signal offline. 
\rev{We control the transmission power of each user so that their reception powers are almost balanced (i.e., within 2dB) and} emulate different SNRs.
For the realistic channel, we consider two, three and four users. 

\textbf{Results.}
Fig.~\ref{fig:BER_SNR}(a) shows the results of different decoders with no relative phase offset.
Given four possible constellation points, two of them (i.e., (+1, -1) and (-1, +1)) are almost overlapped, and are hard to differentiate. It may confuse decoders that need to classify them. We can see that Conv-JDs have an error floor even at high SNRs, and their performances are worse than LDPC-JD decoders.
The key reason is that convolutional decoders rely on joint states, and inseparable constellation points would lead to a wrong survivor path and thus wrong bits. 
In contrast, LDPC-JD aggregates probability for SUM bit decision in the end despite inseparable constellation points during iterations. The poor performance of PSUDs is due to the treatment of the other user's signal as noise.

Fig.~\ref{fig:BER_SNR}(b) shows the results of different decoders with a relative phase offset of $\pi$/2 radian. Four possible constellation points locate on the orthogonal coordinates and equally partition the complex plane into four equal decision regions. In this case, the performance of parallel decoders (PSUDs) is close to joint decoders (JDs), since signals are orthogonal. The performance of Conv-FSJD is close to LDPC-JD, since all constellations are separable for Conv-FSJD. However, due to the probability aggregation capability, LDPC-JD is still better than Conv-FSJD at SNR $\geq$ 6 dB. We also observe that Conv-FSJD outperforms Conv-RSJD since RSJD loses some path information during decoding.

Fig.~\ref{fig:BER_SNR}(c) shows the results of different decoders with a relative phase offset of $\pi$/4 radian. The performance of all decoders gets worse than that with relative phase offset $\pi$/2 radian, and is better than that without relative phase offset. Furthermore, LDPC-JD is less sensitive to the relative phase offset compared with Conv-JD. From Fig.~\ref{fig:BER_SNR}(a), Fig.~\ref{fig:BER_SNR}(b) and Fig.~\ref{fig:BER_SNR}(c), we can observe that phase offset has a strong impact on the SUM bit decoding performance, and LDPC-JD is more robust than Conv-JD against various phase offsets.

Then we show the experiment results on USRPs. In the experiment, \rev{we only consider \emph{Conv-RSJD} for practical convolutional-coded AirComp systems, since \emph{Conv-FSJD} has unacceptable exponential states for more than two users}. We consider \emph{LDPC-JD} for LDPC-coded AirComp. A typical channel in our experiments is shown in Fig.~\ref{fig:BER_SNR_REAL}(b) and Fig.~\ref{fig:BER_SNR_REAL}(c). Fig.~\ref{fig:BER_SNR_REAL}(b) shows the channel amplitudes of all subcarriers for different users. Fluctuating amplitudes in Fig.~\ref{fig:BER_SNR_REAL}(b) indicate a multi-path channel for each user. Fig.~\ref{fig:BER_SNR_REAL}(c) shows the cumulative distribution function of relative phase offset between any two users for subcarriers over all symbols. Due to the long packet duration with 1500 Bytes, the phase offset spreads over the whole plane, and the distribution is almost even although their initial phase offsets may be different. In overall, the realistic channel experiences all possible phase misalignments, and the decoder performance seems to be averaged over fixed phase misalignments as in Fig.~\ref{fig:BER_SNR}.

Fig.~\ref{fig:BER_SNR_REAL}(a) shows the SUM BER results of different decoders under a realistic channel. 
LDPC-JD outperforms Conv-RSJD for all SNRs given the same number of simultaneous transmissions in the realistic channel. The observation is consistent with results in simulated AWGN channels. Furthermore, the decoding performance of LDPC-JD degrades as the number of simultaneous transmissions increases. Given the $10^{-3}$ target for FL applications, two-user LDPC-coded AirComp works at SNR $\geq$ 6 dB, three-user LDPC-coded AirComp works at SNR $\geq$ 10 dB, and four-user LDPC-coded AirComp works at SNR $\geq$ 12 dB.

\subsection{Test Accuracy Performance} \label{sec:eval:fl}

\begin{figure*}[t!]
	\centering
	\subfloat[]{
		\includegraphics[width=2.25in]{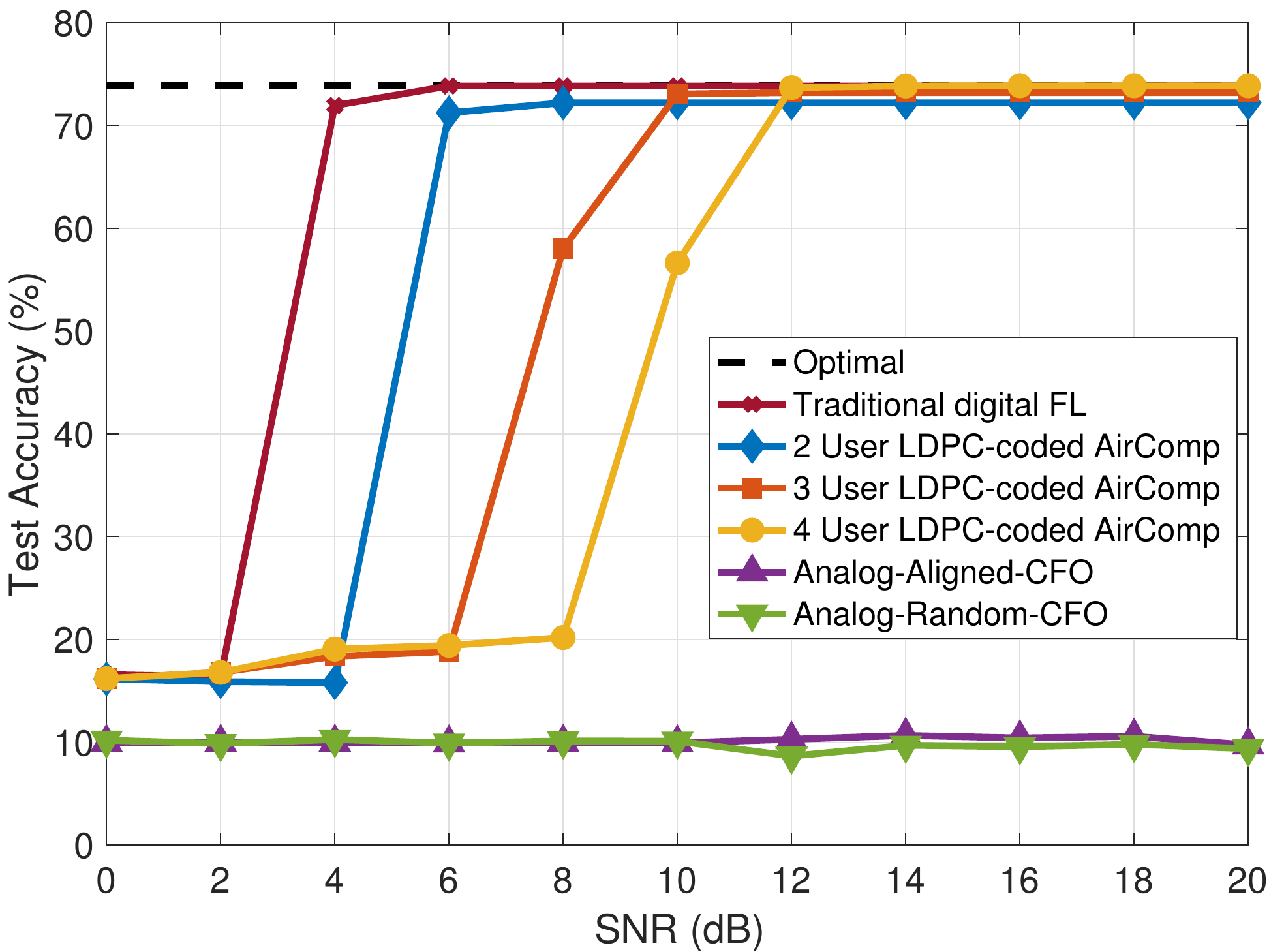} %
		\label{fig:accuracy:shufflenet}}
	\subfloat[]{
		\includegraphics[width=2.25in]{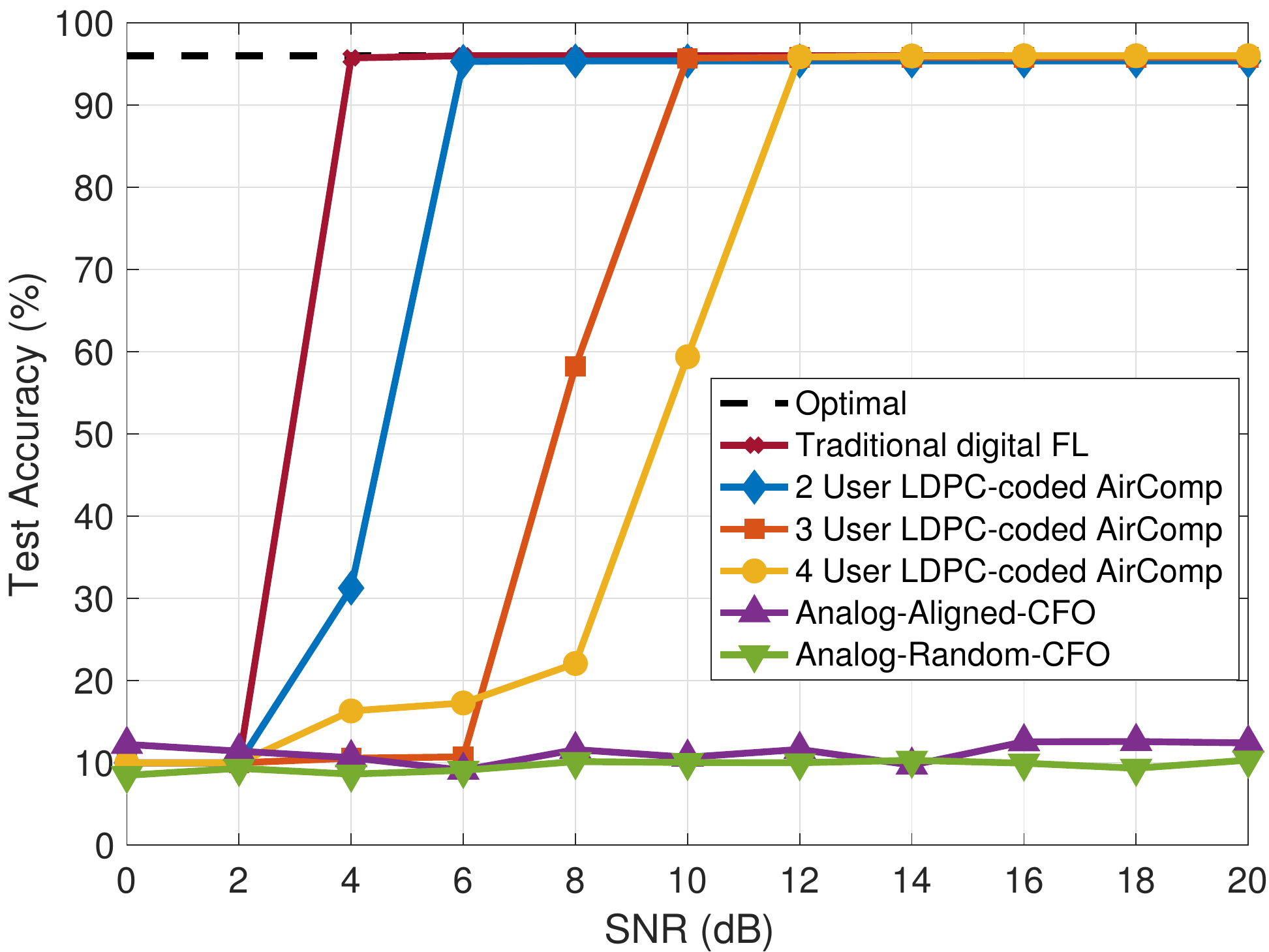} %
		\label{fig:accuracy:mlp}}
	\subfloat[]{
		\includegraphics[width=2.25in]{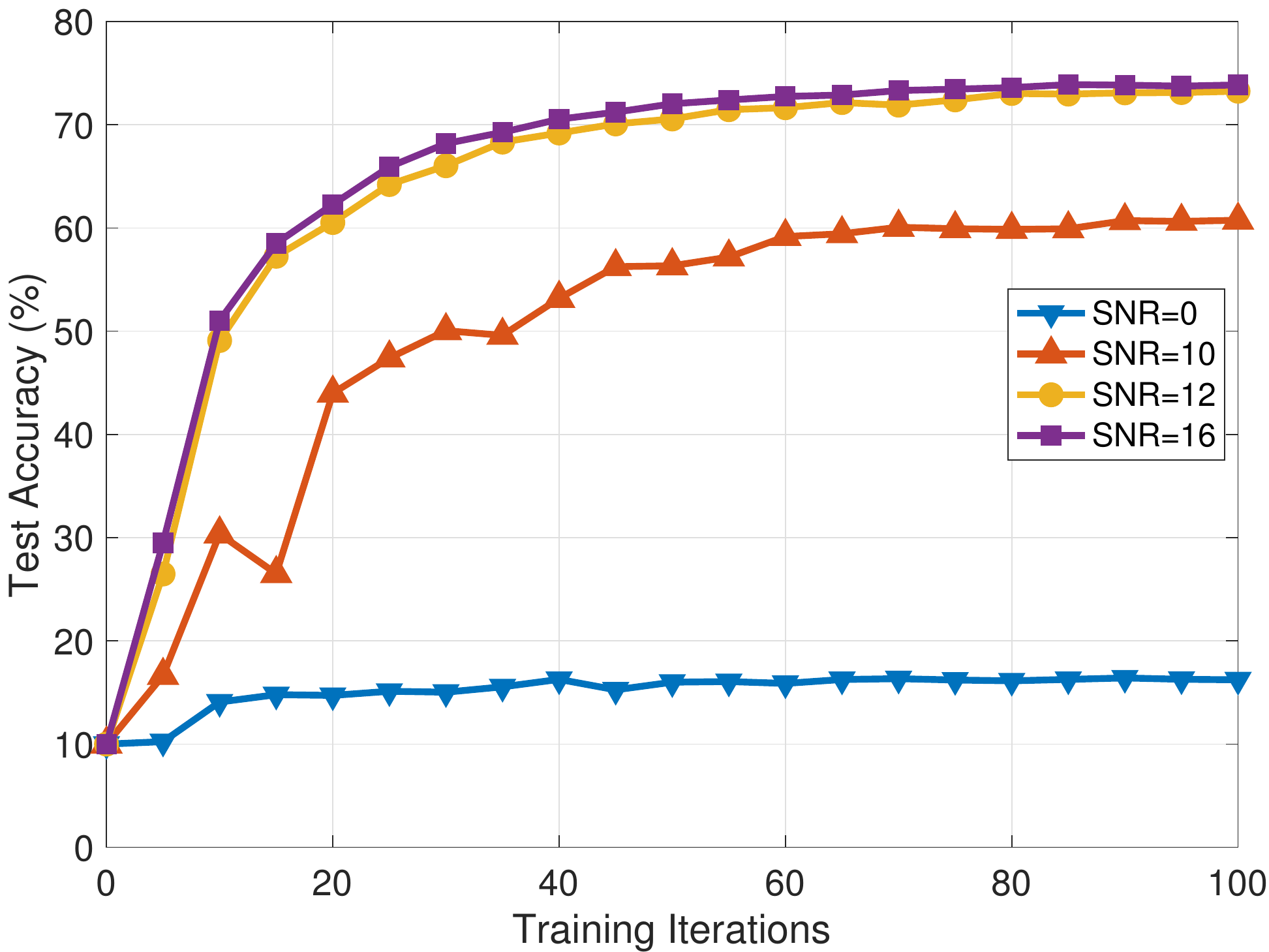}
		\label{fig:accuracy:iter}}
	\caption{Test accuracy performance of different AirComp systems under realistic channel: (a) test accuracy of CIFAR-10 with ShuffleNet-v2 for different AirComp systems; (b) test accuracy of MNIST with MLP for different AirComp systems; (c) test accuracy of CIFAR-10 over training iterations with different SNRs for four-user LDPC-coded AirComp.}
	\label{fig:accuracy}
	\vspace{-2ex}
\end{figure*}

\textbf{FEEL System Setup.}
Our FEEL system consists of a parameter server and \rev{forty} edge devices. They communicate through the wireless medium under various AirComp systems. Digital AirComp systems follow the MAC protocol defined in Section \ref{sec:overview}. We consider two FL applications: CIFAR-10 \cite{Krizhevsky2009} and MNIST \cite{deng2012mnist}. %

For the CIFAR-10 dataset, each edge device uses a convolution neural network (CNN) called ShuffleNet V2 network\cite{Ma2018} to perform classification. CIFAR-10 contains $32\times32$ RGB color images in 10 categories. The size of the CIFAR-10 training dataset is $50,000$ and the size of the testing dataset is $10,000$. The training datasets are distributed in a non-iid manner: 1) we first equally distribute the 40,000 training images to 40 edge devices in random order; 2) the rest 10,000  images are sorted by labels and equally distributed to every device with size 250. %

For the MNIST dataset, each edge device uses an artificial neural network (ANN) called Multiple-layer Perceptron \cite{kubat1999neural} to perform classification. MNIST contains $28\times28$ images of handwriting numbers in 10 categories. The size of the MNIST training dataset is $60,000$ and the size of the testing dataset is $10,000$. The training datasets are also distributed in a non-iid manner.

Every parameter in the neural network is compressed using the probability quantization approach \cite{suresh2017distributed} with 8 bits. In each training iteration, the parameter server randomly selects some edge devices for training. %
We use test accuracy as the metric to evaluate the performance of different AirComp systems. The test accuracy is defined as the prediction accuracy of the global learned model on the test datasets. 

\rev{In the following, we let the number of chosen edge devices be equal to the number of users that the digital AirComp decoder can support (i.e., $P=M$). The test accuracy results for the case $P \neq M$  are similar, so we only present results for the case $P = M$.}

\textbf{\rev{Communication System} Setup.}
We adopt a trace-driven simulation approach to compare the test accuracy performance over SNRs. \rev{We compare digital AirComp with both traditional digital FL system that only allows one user to transmit, and analog AirComp. We adopt the SNR to SUM BER table measured in the realistic channel (including the single-user and multi-user systems). Then, we perform packet-level simulation to simulate the learning process where each frame has 500 data symbols, and the frame may have erroneous SUM bits. For digital AirComp, we only consider the LDPC-JD decoder given its superior performance over the Conv-JD decoder.} 

For analog AirComp \cite{zhu2019broadband}, we assume an AWGN channel with perfect power alignment and time alignment, but with phase misalignment \rev{due to air-channel and CFO}. \rev{For each learning instance, we first generate random air-channels and CFOs for all forty edge devices. Four edge devices are randomly chosen to participate in each training round, and they have different air-channels and CFOs. The final test accuracy results are averaged over 100 instances.}

We compare digital AirComp with the following analog AirComp systems:
\rev{
\begin{enumerate}
    \item \emph{Analog-Aligned-CFO}: It is the analog AirComp with precise precoding at the beginning, and with initial phases in all subcarriers aligned but having rotating phases caused by a random CFO within [-350Hz,~350Hz] over symbols;
    \item \emph{Analog-Random-CFO}: It is the analog Aircomp without precoding, and with initial phases in all subcarriers generated randomly due to air-channel and having rotating phases caused by a random CFO within [-350Hz,~350Hz] over symbols.%
\end{enumerate}
Here the \emph{Analog-Aligned-CFO} AirComp system is to emulate a channel-precoded system with perfect phase alignment at the beginning but with phase misalignment as time goes on. The \emph{Analog-Random-CFO} AirComp system is to emulate a practical system without phase precoding.
}

In analog AirComp systems, the float-valued parameters are directly loaded on subcarriers. We enhance the scheme by allowing each subcarrier to transmit two parameters (in both I and Q planes). To keep the same protocol overhead for time synchronization, we assume the same frame duration as digital AirComp systems.
Since analog AirComp does not need quantization, it leads to less overall data transmission time. To be fair, we allow \rev{an improved analog AirComp system}\footnote{\rev{We also try the analog AirComp system that uses the same duration to train the model with more rounds. However, the results are similar since the performance does not improve after convergence.}} that transmits the same data several times so that the overall transmission duration is the same with digital AirComp (i.e., \rev{16} repeat transmissions for 8-bit quantization in digital AirComp), and \rev{picks the transmission where the aggregated data has the minimum mean squared error compared with the ground truth} as the final result.

\textbf{Results.} Fig.~\ref{fig:accuracy}(a) shows the test accuracy of CIFAR-10 with ShuffleNet-v2  for different systems. First, digital AirComp can realize almost the same performance as the ideal system (i.e., 73\%) at high SNRs. The operating SNR regimes for different digital AirComp systems are consistent with results in Fig.~\ref{fig:BER_SNR_REAL}(a). Second, analog AirComp without accurate phase precoding could not train the model successfully even at high SNRs\footnote{The performance of analog AirComp is lower than the reported performance in our previous conference version \cite{zhao2022broadband}. The reason is that analog AirComp in \cite{zhao2022broadband} assumes the same phase offsets for all subcarriers and all symbols. The phase offsets are generated randomly. Therefore, good phase offsets are possible sometimes, leading to some learning improvement during iterations.}, although initial phases are aligned. The rotating phase offset in a frame induces a large aggregation error. Instead, digital AirComp could overcome the phase asynchrony, and achieve near-optimal learning performance. \rev{Third, the learning performance of traditional digital FL is better than digital AirComp, since it can approach the optimal performance at lower SNR. This phenomenon is as expected, since traditional digital FL can achieve the same BER at lower SNR.} Fig.~\ref{fig:accuracy}(b) shows similar results of MNIST with MLP for different systems.  \rev{The above results reveal the tradeoff between reliability and throughput for various digital FL systems: we can choose a suitable digital FL scheme depending on the SNR ranges that the FEEL system operates.} 

Fig.~\ref{fig:accuracy}(c) shows the test accuracy iteration results of CIFAR-10 for four-user digital AirComp systems in 100 training iterations. We can see that: 1) different SNRs have different convergence rate and results; 2) the learning process converges to the upper bound 73\% test accuracy at SNR $\geq$ 12 dB; 3) the learning process may not converge in a low SNR regime with test accuracy 10\% as the lower bound.
All results show that digital AirComp is more suitable for phase-asynchronous systems.

\subsection{Real-time Performance}
\textbf{Implementation.}
We consider the real-time implementation of LDPC-coded AirComp due to its superior performance over convolutional-coded AirComp. Although the LDPC decoder utilizes the low-complexity BP algorithm for decoding, real-time implementation on the CPU remains a big challenge.
Thanks to the structure of LDPC, parallel processing can be applied on check nodes and variable nodes in each iteration to accelerate the decoding. Moreover, parallel processing can also be applied on independent blocks for a long frame. Some previous work \cite{Tarver2021gpu} has demonstrated superior decoding performance of single-user LDPC decoder on the graphics processing unit (GPU) hardware. Therefore, we also adopt GPU to implement LDPC Jt-CDA. 

The digital AirComp receiver is implemented on a Ubuntu 18.04.6 LTS server with Intel Xeon Gold 5122 CPUs (four cores@3.6GHz), 64GB DDR4 DRAM, and an Nvidia GeForce RTX 2080 Ti GPU with CUDA 11.4. The LDPC decoding function is written in CUDA C, and is compiled to a dynamic library using \emph{nvcc}. We modify an open-source single-user LDPC GPU decoder \cite{ldpcgpu} to realize the multi-user joint decoder defined in Section~\ref{sec:ldpc}. There are 40 iterations during the LDPC decoding.
Then we write a GNU Radio block in Python that loads the library during initialization and calls the decoding function when soft demodulation inputs of a complete frame are ready. The other blocks except the demodulation and decoding blocks are the same as in a real-time PNC system \cite{you2016reliable}.

\textbf{Setup.}
We first measure the \emph{channel decoding time} that decodes the SUM bits using LDPC Jt-CDA. The block is executed for a complete frame, and thus we can add two timers in the decoding block to measure the processing time. Then we try to measure the \emph{frame decoding time} that captures the processing time of all blocks in the receiver. However, it is challenging to measure the time in GNU Radio, since GNU Radio is a pipeline framework that runs in multiple threads, and there are queuing delays among blocks. If we simply compute the time gap between the time a frame is detected and the time bits are decoded, the frame decoding time will be over-estimated, because the time contains the queuing time. To character the processing time of the receiver, we use the \emph{inter-packet interval} (IPI) in the transmitter to approximate the frame decoding time. In particular, we 
tune the IPI, and find the minimal IPI that the receiver is still stable. Here stable means that the inter-packet decoded time gap is equal to the inter-packet arrival time gap (i.e., IPI). %
The method has been adopted in measuring the real-time performance of a PNC system on USRP and GNU Radio in \cite{you2016reliable}.

We also measure the \emph{network throughput} in bits per second considering the protocol overhead.
To improve the network throughput, we extend the designed MAC protocol with a burst transmission mode \cite{Lu2013}, where a trigger frame initiates a burst transmission consisting of multiple frames from each user. The transmission gap is equal to the IPI. The number of frames in a burst is chosen so that the last frame remains time synchronized despite clock drifts in each user. Note that the burst MAC reduces the synchronization overhead, since one trigger frame synchronizes multiple data frames. %
To take the number of concurrent transmissions into account, we define the network throughput as the product of the measured throughput in the edge server and the number of concurrent transmissions.

We measure and fix the IPI (i.e., frame decoding time) beforehand for different numbers of simultaneous transmissions. We also measure and fix the burst duration of 1s. Then we measure the throughput 30 times with each measurement 500s, and present the average result. Meanwhile, we measure the channel decoding time for each frame, and present the average result with standard deviation over all measurement. In each measurement, there are 25,000, 16,500, 12,500 frames for two-user, three-user, and four-user digital AirComp, respectively.

\textbf{Results.}
Table~\ref{table:rtime} shows the overall real-time performance results.
The Jt-CDA time is almost linear to the number of simultaneous transmissions. It is consistent with Algorithm~\ref{algo:bp} since the complexity increases with the message vector length. The frame decoding time is larger than the Jt-CDA time, since Jt-CDA is just only one step of the receiver. 
The network throughput is about 1.19 Mbps for two-user LDPC-coded AirComp, and 1.16 Mbps for four-user LDPC-coded AirComp.  Given that the frame duration is around 4ms, and the physical throughput is 3Mbps for our system (i.e., around 500 data symbols, 10MHz bandwidth, BPSK and 1/2 coding rate), the achievable throughput almost reaches the upper bound for the two-user LDPC-coded AirComp (i.e., 4/20 $\times$ 3Mbps  $\times$ 2 = 1.2Mbps), meaning negligible protocol overhead.
Although the achievable throughput is lower than the physical throughput, we note that the bottleneck is on the Jt-CDA block, and further optimization is possible with more powerful GPU.

\begin{table}[t]
	\centering
	\caption{Real-time performance of  the LDPC-coded AirComp system with two, three, and four concurrent users. CD time means channel decoding (i.e., Jt-CDA) time, and FD time means frame decoding time (i.e., IPI).}
	\setlength{\tabcolsep}{3.5mm}{
		\begin{tabular}{|c|c|c|c|}
			\hline \textbf{} & \textbf{CD Time} & \textbf{FD Time} & \textbf{Throughput}\\
			\hline Two-User & 17.2$\pm$1.6ms & 20ms & 1.19Mbps \\ %
			\hline Three-User & 24.5$\pm$2.0ms & 30ms & 1.18Mbps \\ %
			\hline Four-User & 38.9$\pm$3.1ms & 40ms & 1.16Mbps \\ %
			\hline
		\end{tabular}}
		\label{table:rtime}
\end{table}

\section{Discussion} \label{sec:discussion}

\rev{
\textbf{Real-time FEEL.} We have presented a real-time digital AirComp system for FEEL. However, it is still far from a complete and real-time FEEL system. Although communication is a bottleneck that restricts the development of FEEL \cite{McMahan2017}, device heterogeneity with different local training time and data heterogeneity with different data distribution also have a strong impact on the FEEL performance, and thus user selection that considers device and data heterogeneity has been widely studied in FEEL. Moreover, there exists communication heterogeneity in FEEL with digital AirComp: edge devices have heterogeneous channels and CFOs, and different combinations of simultaneous transmissions have different communication performances. Therefore, one important direction of future work is to investigate the user selection strategy to improve the overall learning performance, and develop a real-time FEEL system.

\textbf{Joint Source-Channel Coding Design.} We believe that channel codes are an indispensable part of any communication system supporting wireless FL, since bit errors are inevitable due to noise, frequency-selective fading and adverse phase offsets.
However, as shown in Figure \ref{fl} in Section \ref{sec:motivation:convergence}, FL applications may not require a high degree of reliability, and may allow highly-compressed source codes for communication. The joint design of source codes and channel codes is an interesting direction for future work.
}

\section{Related Work} \label{sec:related}

\textbf{AirComp Systems.}
Over-the-air computation in multiple-access channels is first studied from an information-theoretic perspective \cite{gastpar2003source, nazer2007computation}. They show that joint communication and computation can be more efficient than separating communication from computation in terms of function computation rate. Recently, many AirComp systems \cite{amiri2019over, zhu2019broadband, xing2020decentralized, yang2020federated, amiri2020machine, zhu2020one, sery2021over, wang2021federated, shao2021denoising, wang2022interference, zou2022knowledge} for FEEL have been proposed. However, all of them are analog AirComp systems, and require accurate power and phase synchronization.

Abari, Rahul and Katabi \cite{abari2016over} try to realize analog AirComp for the first time. There are some analyses in \cite{abari2016over}, but the real system is not demonstrated. Guo et al., \cite{guo2021over} provide the first implementation of two-user analog AirComp. It leverages dedicated frame/protocol design and accurate timing tracking to compensate for phase misalignment. 
Sahin \cite{csahin2022demonstration} provides a new five-user analog AirComp implementation for SignSGD-based FL. Different from previous analog AirComp systems, it uses two subcarriers for the positive vote and the negative vote, and compares their energy for decisions based on the majority vote rule. 
\rev{The positive results are analyzed using statistical channel, and some preliminary results are presented via SDR experiments. In contrast, we take another way to deal with phase asynchrony following the standard physical layer,}
and demonstrate a four-user digital AirComp system.

Some analog AirComp systems are proposed to handle phase-misaligned transmissions. Goldenbaum et al., \cite{goldenbaum2013robust, kortke2014analog} design and implement an analog-modulated AirComp system that only needs coarse symbol-level synchronization, and leverages direct spread spectrum sequences with random phases for modulation. Shao et al., \cite{shao2021federated,shao2022bayesian} leverages oversampling and sum-product ML estimators to estimate the arithmetic sum under symbol misalignment and phase misalignment. However, their designs only work for time-domain narrowband systems, and cannot directly apply to the frequency-domain OFDM broadband system.

\textbf{Network MIMO Systems.} Network MIMO systems are envisioned to improve network capacity, where distributed APs in different locations form a virtual big AP for simultaneous transmissions and receptions to several users (i.e., distributed MU-MIMO). The most important requirement is to maintain frequency synchronization among distributed APs during transmissions. Otherwise, CFOs between APs and users rotate the channel phases, and invalidate the orthogonality of inter-user signals which is achieved by the precoding matrix computed from the initial channel.

Many solutions have been proposed.
However, all these solutions require extra resources and dedicated protocol design. Out-of-band radio \cite{balan2013airsync}, out-of-band power line infrastructure \cite{yenamandra2014vidyut}, out-of-band Ethernet infrastructure \cite{rahul2012jmb, wang2017dcap}, and dedicated hardware \cite{abari2015airshare} are required to realize frequency synchronization. Moreover, receivers need to 
provide feedback information on the initial subcarrier phases for interference-free transmissions. Phase-aligned transmissions are not easy for random-access WiFi networks if low overhead is desired. Therefore, we target low-cost multi-user OFDM systems, and design signal processing algorithms for digital AirComp.

\section{Conclusion}\label{sec:conclusion}

This paper present \rev{the first digital AirComp system using OFDM modulation} for phase-asynchronous FEEL systems. With a simple multi-user non-orthogonal channel access protocol with negligible overhead, our system leverages joint channel decoding and aggregation decoders tailored for convolutional and LDPC codes to overcome phase asynchrony. We implement a real-time digital AirComp system with up to four users on a software-defined radio platform. 
Trace-driven simulation results show that digital AirComp can achieve near-optimal learning performance under practical phase-asynchronous scenarios, while analog AirComp fails to do so even with high SNRs. 

Moving forward, our digital AirComp system can be extended in the following directions: 
\rev{a) SUM bit-optimal decoder design for Turbo-coded AirComp}, b) joint source-channel coding designs to directly improve the application performance, and c) user selection in FEEL with digital AirComp for improved network performance.

\appendices

\ignore{
\section{Relationship between time domain channel and frequency domain channel}
This section shows the relationship between the time-domain channel and the frequency-domain channel. Let $h^u (t)=[h_1^u,\dots,h_L^u]$ be a complex vector, representing the time-domain channel of user $u$ with $L$ paths, where the $l$-th path has delay $\tau_l^u$. Let $H^u[k]$ be the frequency-domain channel of the $k$-th subcarrier. Assume the demodulation window is aligned with the first path. Then we have
\begin{equation}
    H^u[k]=\sum_{l=1}^L h_l^u e^{j\frac{2\pi\tau_l^u k}{N}},
\end{equation}
where $N$ is the FFT length. According to the above formula, the multi-path channel introduces phase misalignment even with the cyclix prefix (CP) design in OFDM since $h_l^u$ and $\tau_l^u$ are different from different users.

If we further consider time offset (TO), then we have
\begin{equation}
    H^u[k]=\sum_{l=1}^L h_l^u e^{j\frac{2\pi(\tau^u+\tau_l^u)k}{N}},
\end{equation}
where $\tau^u$ is the TO of user $u$. According to the above formula, TO introduces additional phase misalignment related to the TO value. Note that the above phase misalignment is fixed for all symbols. However, if we consider carrier frequency offset (CFO), as shown in Eq. (2), the phase misalignment rotates over symbols, making phase-aligned transmissions more challenging.
}

\rev{
\section{Relationship between BER and SUM BER} \label{appendix:prob}
For the multi-user digital AirComp system, the SUM bit is the arithmetic summation of all users’ original bits, and thus the SUM bit error is different from the traditional bit error. In particular, for the $n$-user digital AirComp system, the corresponding mapping relationship of the original bits and the SUM bit is given as follows
$$\underbrace{{\{ 0,1\} } \times \cdots \times {\{ 0,1\} }}_n \to {\{ {\rm{0,1,\cdots,n}}\} }$$
with the mapping function from size $2^n$ to size $\left(n+1\right)$. The unequal size makes them different.

Let the traditional source bit error rate (BER) be $\alpha$, and the SUM bit error rate (SUM BER) be $\beta$. This appendix analyzes their relationship. We assume the error in each bit is independent (channel codes can be used to make the assumption valid).

Take the two-user digital AirComp system for example. There are four kinds of two-users’ bit sequences $\{00, 01, 10, 11\}$, while there are three kinds of  SUM bits $\{0, 1, 2\}$. If the original bit sequence is 00 or 11, any bit error causes the SUM bit different. Therefore, the SUM bit error probability is given by $\beta=1-\left(1-\alpha\right)^2$. If the original bit sequence is 01 or 10, one-bit error causes the SUM bit different. Thus, the SUM bit error probability is given by $\beta=1-\left(1-\alpha\right)^2-\alpha^2$. Given the equal probability of sequences $\{00, 01, 10, 11\}$ (i.e., 1/4), the SUM bit error rate is given by $1-\left(1-\alpha\right)^2-\frac{\alpha^2}{2}$. Obviously, $\beta<1-\left(1-\alpha\right)^2$. Given $\alpha$ is typically much smaller than ${10}^{-3}$ in a communication system, $\beta>\alpha$ (it holds for $\alpha<2/3$).

For the $n$-user digital AirComp system, we first consider two special multi-user’ bit sequence with length $n$ $\{0\ldots0, 1\ldots1\}$, their SUM BER is given by 
$$\beta_{\underbrace{0\ldots0}_n} =  \beta_{\underbrace{1\ldots1}_n} = 1-\left(1-\alpha\right)^n.$$ 
For other multi-user’ bit sequences, we consider a particular sequence $s$ with $n_0$ zeros and $n-n_0$ ones ($0 < n_0 < n$). The SUM BER is given by 
\begin{equation}
    \beta_s\ =\ 1-\left(1-\alpha\right)^n-\sum_{e=1}^{min\{n_0,n-n_0\}}\alpha^{2e}.
\end{equation} 
Then the overall SUM BER is given by 
\begin{equation} \label{eq:sum-ber}
\beta = 1-\left(1-\alpha\right)^n - \frac{1}{2^n} \sum_{n_0=1}^{n-1} {n \choose n_0} \sum_{e=1}^{min\{n_0,n-n_0\}} \alpha^{2e}.    
\end{equation}

That is, BER $\alpha$ and SUM BER $\beta$ are one-to-one correspondence.
Obviously, $\beta < 1-\left(1-\alpha\right)^n$. Then we are to prove $\beta > \alpha$ for $0 < \alpha<{10}^{-3}$. 

Given $min\{n_0,n-n_0\} \leq n/2$, we have
\begin{equation}
    \sum_{e=1}^{min\{n_0,n-n_0\}}\alpha^{2e} \leq \sum_{e=1}^{n/2}\alpha^{2e} = \alpha^2 \frac{1-\alpha^n}{1-\alpha^2} < \alpha.
\end{equation}
The last inequality holds since $\alpha+\alpha^2 < 1+\alpha^{n+1}$ for $0 < \alpha < 10^{-3}$. Then Eq.~(\ref{eq:sum-ber}) becomes
\begin{equation} 
\beta > 1-\left(1-\alpha\right)^n - \alpha.    
\end{equation}

Let $f(n)=1-(1-\alpha)^n - 2\alpha$. Then we are to prove $f(n)>0$ for $n \geq 3$  by induction, where $0<\alpha<{10}^{-3}$.
\begin{itemize}
    \item For $n=3$, $f(3)=1-(1-\alpha)^3 - 2\alpha=\alpha^3 - 3 \alpha^2 + \alpha = \alpha(\alpha^2 - 3\alpha + 1)$. Let $g(\alpha)=\alpha^2 - 3\alpha + 1$. It is easy to check that $g(\alpha)$ is a monotonically decreasing function for $\alpha \in (0, 10^{-3})$. Thus, $g(\alpha) > 10^{-6}-3\cdot10^{-3}+1 > 0$, and $f(3)>0$ for $\alpha \in (0, 10^{-3})$.
    \item We assume $f(n) = 1-(1-\alpha)^n - 2\alpha > 0$ holds.
    \item We consider $f(n+1)$:
    \begin{equation}
    \begin{aligned}
        f(n+1) &= 1-(1-\alpha)^{n+1} - 2\alpha \\
        &= (1-2\alpha)-(1-\alpha)^{n+1} \\
        &> (1-\alpha)^{n}-(1-\alpha)^{n+1} > 0, \\
    \end{aligned}
    \end{equation}
    where $(1-2\alpha) > (1-\alpha)^{n}$ holds due to the assumption for $n$.
\end{itemize}

Overall, SUM BER is larger than BER (i.e., $\beta > \alpha$), meaning the SUM bit indeed differs from the traditional bit.
}

\section{Proofs of Convergence Analysis} \label{appendix:proof}

\subsection{Proof of Lemma 1}
\begin{proof}
First, 
\begin{equation}
    \begin{aligned}
        \label{23}
        \mathbb{E}(B(\Delta w_n(t)))&= \mathbb{E}[Q(\Delta w_n(t))+X(\Delta w_n(t))] \\
        &=\mathbb{E}[Q(\Delta w_n(t))]+\mathbb{E}[X(\Delta w_n(t))] \\
        &\overset{(a)}{=}\mathbb{E}[Q(\Delta w_n(t))] \\
        &\overset{(b)}{=}\Delta w_n(t),
    \end{aligned}
\end{equation}
where (a) holds because of Eq.~(13) in Lemma 1, and (b) holds due to Lemma 1 in \cite{bouzinis2022wireless}.
Then,
\begin{equation}
    \begin{aligned}
        &\mathbb{E}[\Vert B(\Delta w_n(t))-\Delta w_n(t)\Vert_2^2] \\
        &=\mathbb{E}[\Vert Q(\Delta w_n(t))+X(\Delta w_n(t))-\Delta w_n(t)\Vert_2^2] \\
        &=\mathbb{E}[\Vert Q(\Delta w_n(t))-\Delta w_n(t)\Vert_2^2]+\mathbb{E}[\Vert X(\Delta w_n(t))\Vert_2^2] \\
        &+2\mathbb{E}[\left \langle Q(\Delta w_n(t))-\Delta w_n(t),X(\Delta w_n(t))\right \rangle ] \\
        &\overset{(c)}{=}\mathbb{E}[\Vert Q(\Delta w_n(t))-\Delta w_n(t)\Vert_2^2]+\mathbb{E}[\Vert X(\Delta w_n(t))\Vert_2^2] \\
        &\overset{(d)}{\leq}J_n^2(t)+K_n(t),
    \end{aligned}
\end{equation}
where (c) holds because 
$$\mathbb{E}[\left \langle Q(\Delta w_n(t))-\Delta w_n(t),X(\Delta w_n(t))\right \rangle ]=0,$$ 
which is due to (b) in Eq.~(\ref{23}).

We divide it into two parts to prove that (d) holds.
First, we have the following inequality in \cite{bouzinis2022wireless}:
\begin{equation}
    \label{26}
    \mathbb{E}[\Vert Q(\Delta w_n(t))-\Delta w_n(t)\Vert_2^2]\leq \frac{\delta_n^2(t)}{(2^{B_n(t)}-1)^2}=J_n^2(t).
\end{equation}
Second, we can have:
\begin{equation}
    \label{27}
    \begin{aligned}
        &\mathbb{E}[\Vert X(\Delta w_n(t))\Vert_2^2] \\
        &=\frac{\alpha}{2}\sum_{i=0}^{B_n(t)-1}\left((\frac{4\delta_n(t)^2}{d(2^{B_n(t)}-1)}\times2^i\times1)^2+\right.\\
        &\left.\quad(\frac{4\delta_n(t)^2}{d(2^{B_n(t)}-1)}\times2^i\times(-1))^2\right)\\
        &=\alpha(\frac{4\delta_n(t)^2}{d(2^{B_n(t)}-1)})^2\sum_{i=0}^{B_n(t)-1}2^{2i}\\
        &=\alpha(\frac{4\delta_n(t)^2}{d(2^{B_n(t)}-1)})^2\times\frac{1-4^{B_n^2(t)}}{1-4}\\
        &\triangleq K_n(t).
    \end{aligned}
\end{equation}
The proof is completed.
\end{proof}

\subsection{Proof of Theorem 1}
\begin{proof}
We follow the method in \cite{bouzinis2022wireless}. We first introduce the lossless model at the ($t$+1)-th communication round as
\begin{equation}\label{eq24}
    \hat{w}(t+1)=w(t)+\frac{1}{N}\sum_{n=1}^Np_n\Delta w_n(t).
\end{equation}
With Eq.~(\ref{eq24}), we have
\begin{equation}
    \label{eq26}
    \begin{aligned}
        \Vert w(&t+1)-w^* \Vert_2^2 \\
        &=\Vert w(t+1)-\hat{w}(t+1)+\hat{w}(t+1)-w^* \Vert_2^2 \\
        &=\Vert w(t+1)-\hat{w}(t+1)\Vert_2^2 + \Vert \hat{w}(t+1)-w^* \Vert_2^2 \\
        &+2\left \langle w(t+1)-\hat{w}(t+1),\hat{w}(t+1)-w^* \right \rangle.
    \end{aligned}
\end{equation}
We split the equation into three parts and analyze their property separately. The second part and the third part are the same with \cite{bouzinis2022wireless}. So we only prove the first part.

For the first part, we can have
\begin{equation}
    \begin{aligned}
        \mathbb{E}&\left [ \Vert w(t+1)-\hat{w}(t+1)\Vert_2^2 \right ] \\
        &=\mathbb{E}\left [ \Vert \sum_{n=1}^Np_n(Q(\Delta w_n(t))+X(\Delta w_n(t))-\Delta w_n(t))\Vert_2^2 \right ] \\
        &\leq \sum_{n=1}^Np_n\mathbb{E}\left [ \Vert Q(\Delta w_n(t))+X(\Delta w_n(t))-\Delta w_n(t)\Vert_2^2\right ] \\
        &\leq \sum_{n=1}^Np_n(J_n^2(t)+K_n(t)).
    \end{aligned}
\end{equation}
The first inequality follows from the convexity of $\Vert · \Vert_2^2$ and $\sum_{n=1}^Np_n=1$, and the second inequality follows from the Lemma 1.

\ignore{
For the middle part, we can have
\begin{equation}
    \begin{aligned}
        \mathbb{E}\left [\hat{w}(t+1)-w^* \right ] \leq -\mu\eta(t)\mathbb{E}\left [w(t)-w^* \right ]+\eta^2(t)U
    \end{aligned}
\end{equation}
where
\begin{equation}
    \begin{aligned}
        U=\tau^2\sum\sigma_n^2+\tau G^2+2L\tau^2\Gamma+(\mu+2)\frac{\tau(\tau-1)(2\tau-1)}{6}G^2 
    \end{aligned}
\end{equation}
which is proved by \cite{bouzinis2022wireless}.

For the last part, we can have
\begin{equation}
    \begin{aligned}
        \mathbb{E}\left [2\left \langle w(t+1)-\hat{w}(t+1),\hat{w}(t+1)-w^* \right \rangle \right ]=0,
    \end{aligned}
\end{equation}
which can be proved with $\mathbb{E}\left [Q(\Delta w_n(t)) \right] = \Delta w_n(t)$
and
\begin{equation}
    \begin{aligned}
        &w(t+1)-\hat{w}(t+1) \\
        &= \sum_{n=1}^Np_n(Q(\Delta w_n(t))+X(\Delta w_n(t))-\Delta w_n(t)).
    \end{aligned}
\end{equation}
}

Based on the above analysis, we can rewrite Eq.~(\ref{eq26}) as
\begin{equation}
    \begin{aligned}
        \mathbb{E}\left[\Vert w(t+1)-w^* \Vert_2^2 \right]&\leq (1-\eta(t)\mu)\mathbb{E}\left[\Vert w(t)-w^* \Vert_2^2 \right]\\
        &+\eta(t)^2U+\sum_{n=1}^Np_n(J_n^2(t)+K_n(t)).
    \end{aligned}
\end{equation}
Then following the same proof in \cite{bouzinis2022wireless}, Theorem 1 is proved.
\end{proof}

\rev{
\begin{figure}[t!]
	\centering
	\setlength{\abovecaptionskip}{0.cm}
	\includegraphics [scale=0.4,trim=0 0 0 0]{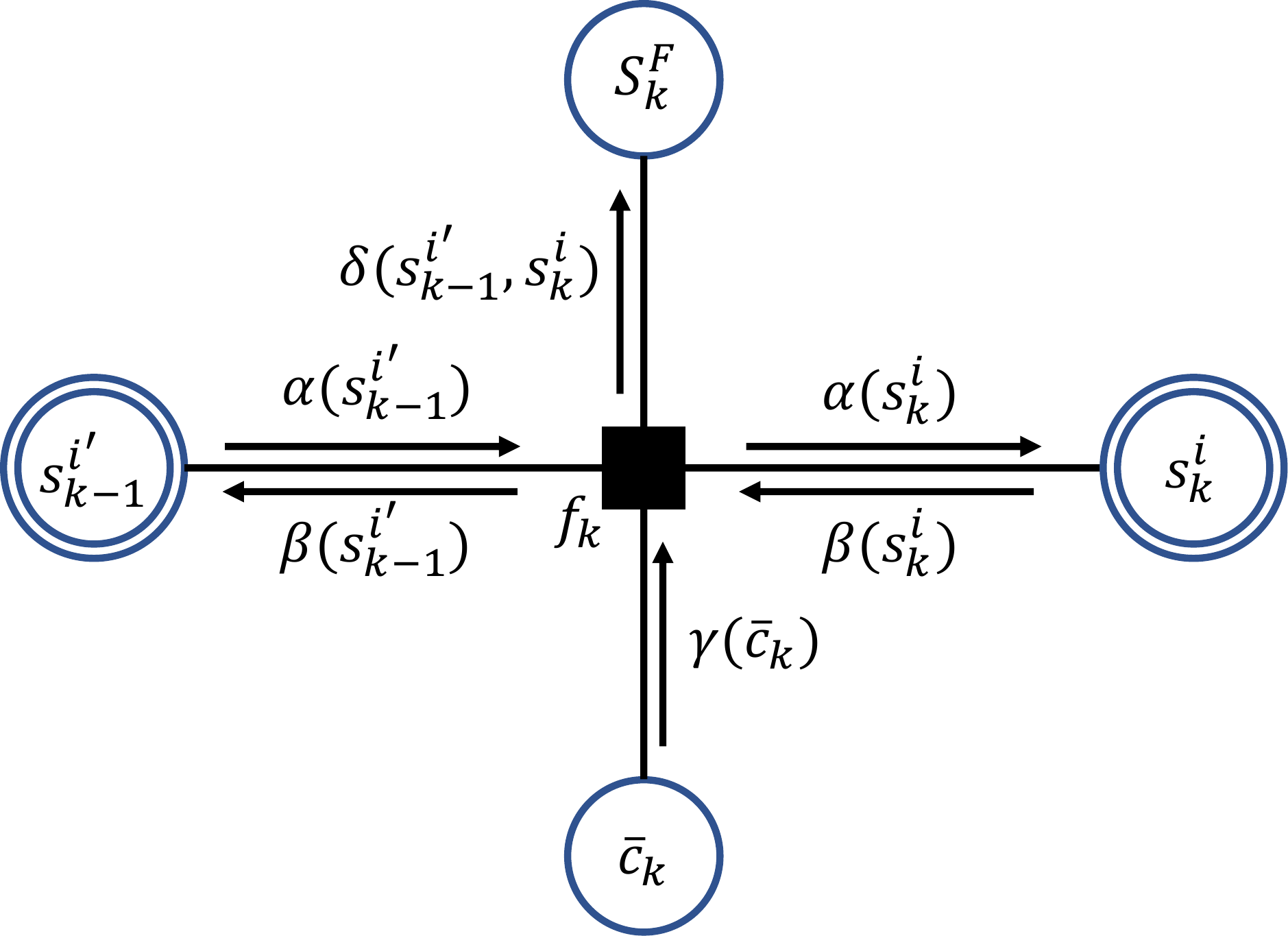}
	\caption{The passed messages around a factor node during the operation of the BCJR algorithm.}
	\label{fig:bcjr}
\end{figure}

\section{Bit-optimal Convolutional Decoder}
The goal of bit-optimal convolutional decoder is to find the maximum-likelihood (ML) SUM bit.
The Bahl-Cocke-Jelinek-Raviv (BCJR) algorithm is the widely used algorithm for bit-optimal decoding. It is a probabilistic iterative decoding algorithm that utilizes message passing to compute the most probable original message sequence based on the received sequence. The core idea of the BCJR algorithm is to use a recursive algorithm to calculate the state transition metrics between each state by computing the branch metrics, and then calculate the probabilities of each state associated with the past, the present and the future of received signal. We modify the BCJR algorithm to support SUM bit decoding. In particular, a posterior probability (APP) of the SUM bits is computed based on these three probabilities.

\begin{figure*}[t]
	\centering
	\subfloat[]{
		\includegraphics[width=2.25in]{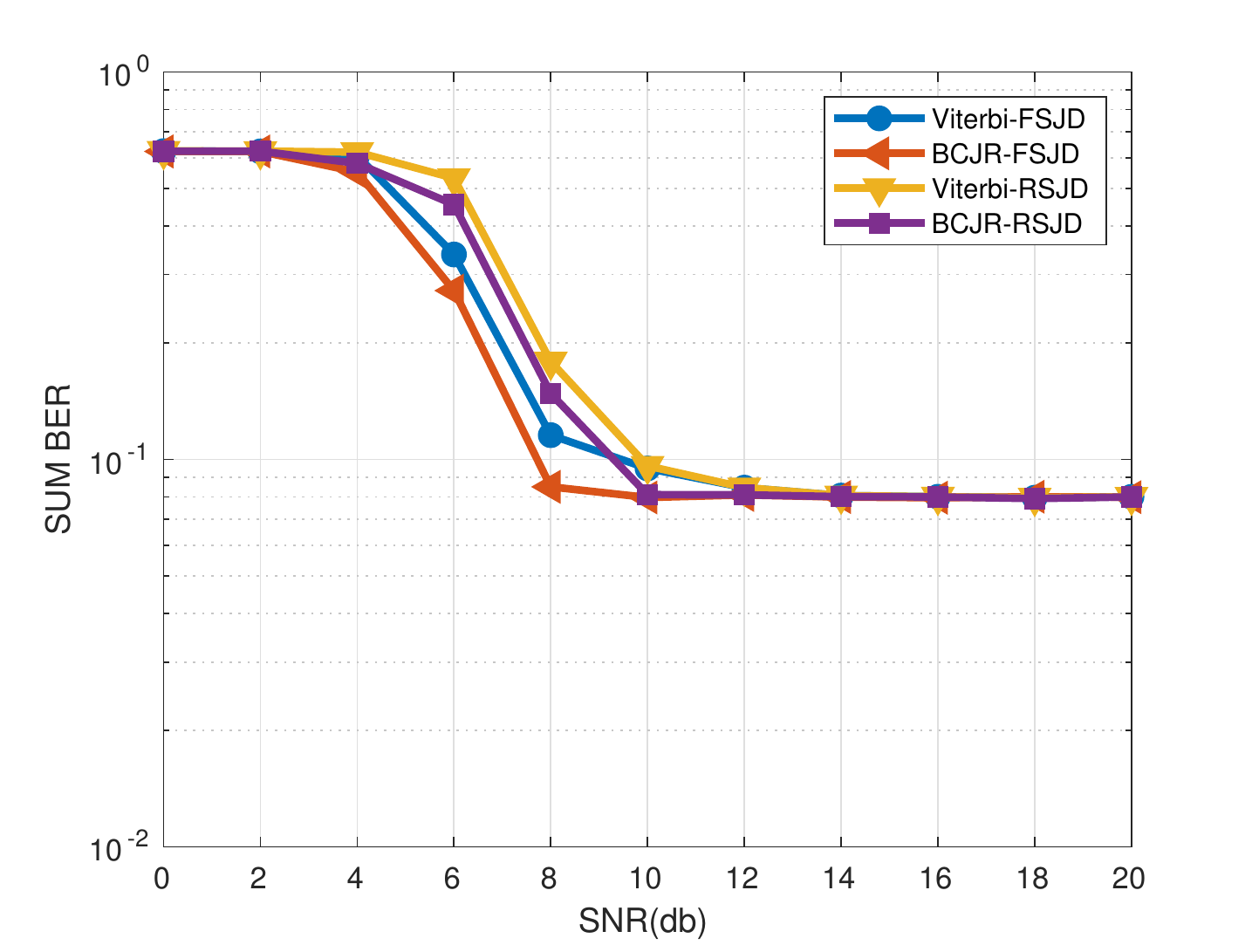} %
		\label{fig:bcjr_00}}
	\subfloat[]{
		\includegraphics[width=2.25in]{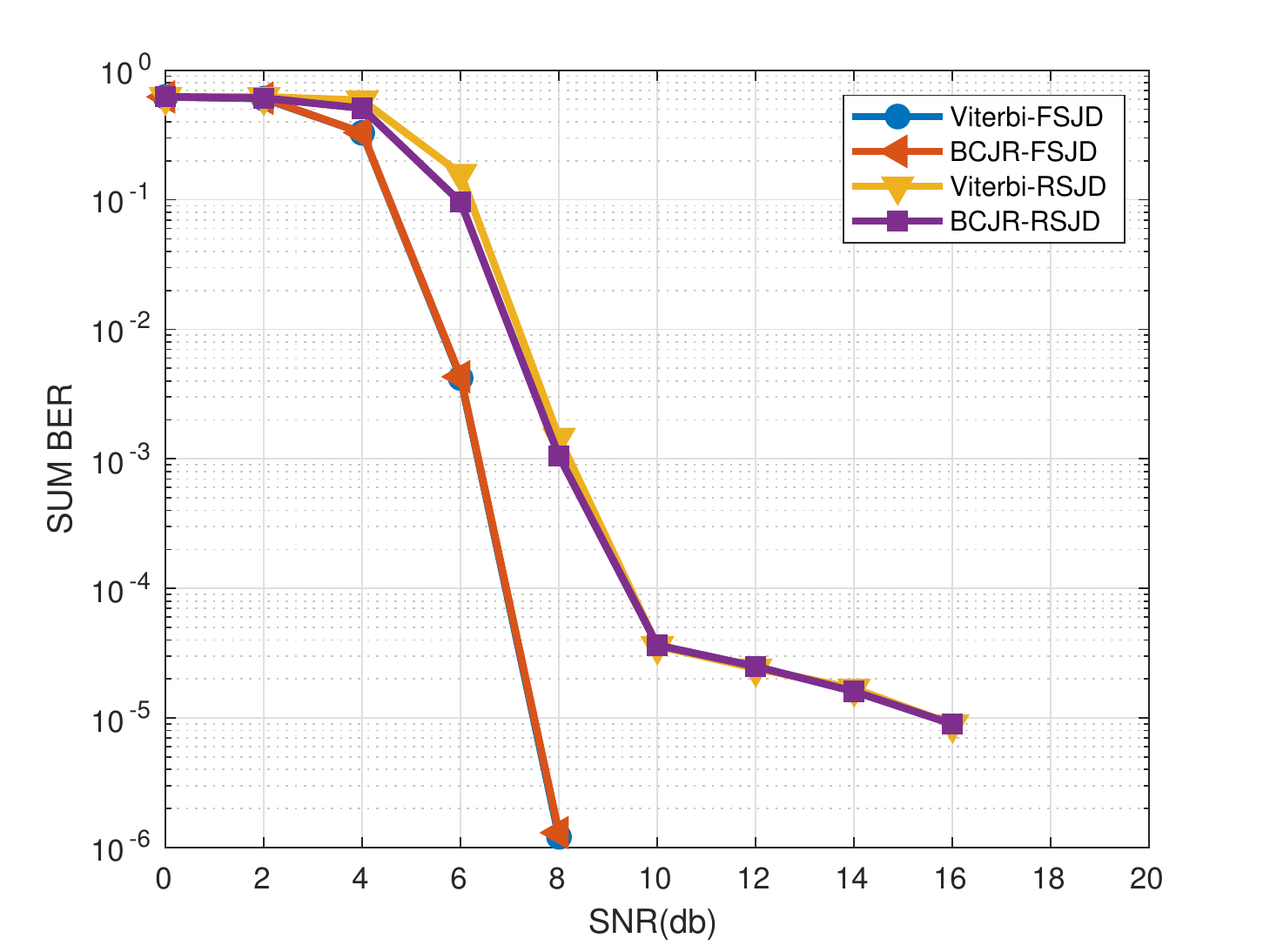} %
		\label{fig:bcjr_01}}
	\subfloat[]{
		\includegraphics[width=2.25in]{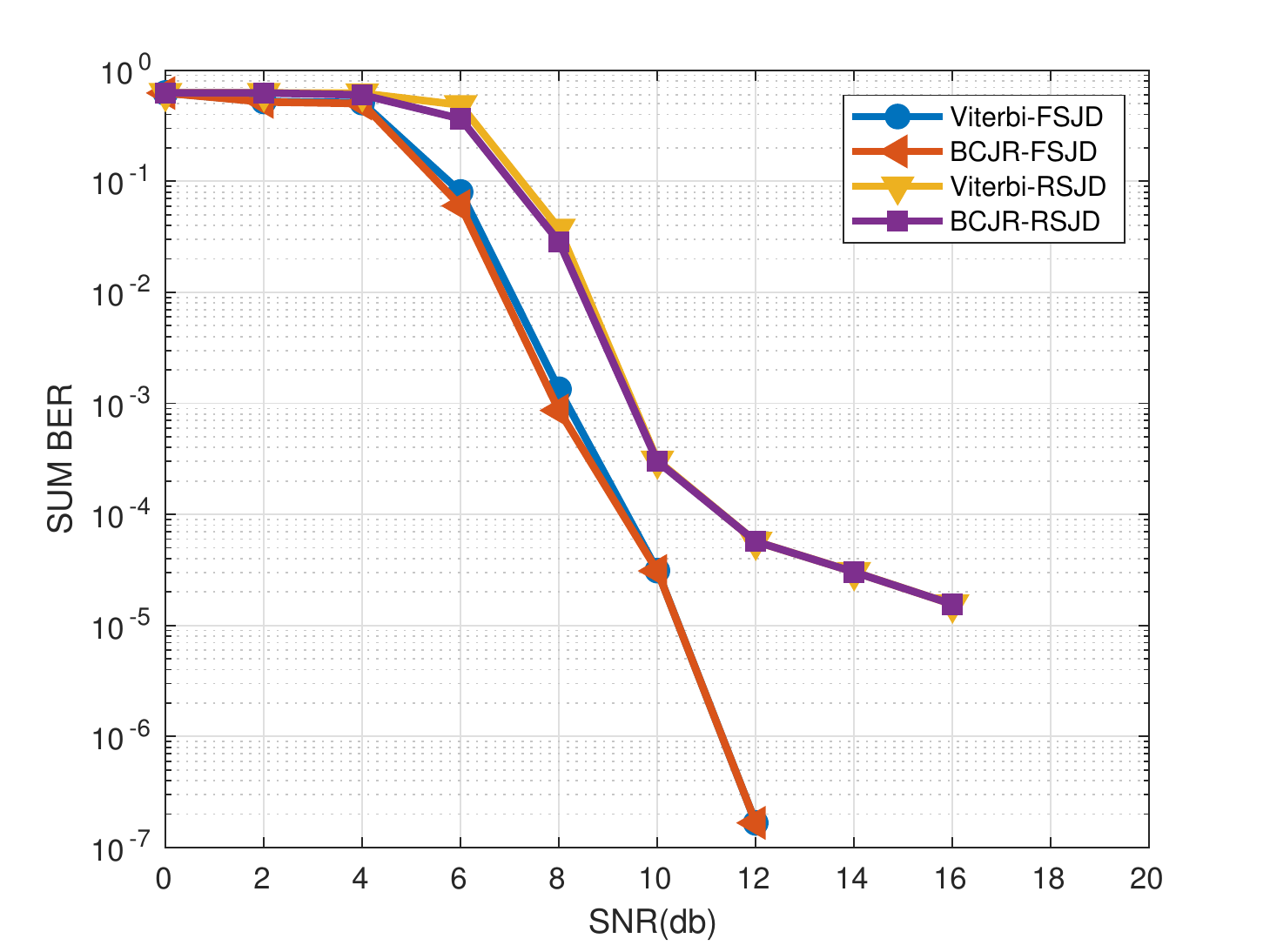} %
		\label{fig:bcjr_0q}}
	\caption{SUM BER versus BER results for the BCJR and Viterbi algorithms under various phase settings for two-user convolutional-coded AirComp: (a) perfectly aligned channel phase; (b) relative channel phase offset of $\pi$/2 radian; (c) relative channel phase offset of $\pi$/4 radian.}
	\label{fig:bcjr_viterbi}
	\vspace{-2ex}
\end{figure*}

\subsection{Algorithm Design}
Similar to the design for codeword-optimal convolutional decoder, we adopt the full-state joint trellis or the reduced-state joint trellis for two-user digital AirComp for bit-optimal convolutional decoder. However, for more than two users, we adopt the reduced-state joint trellis due to the exponential number of joint states.

Bit-optimal convolutional decoder passes probability message between states. Fig.~\ref{fig:bcjr} shows the message transmission graph between two states, where $s_{k-1}^{i'}$ represents the $i'$-th state in the $(k-1)$-th stage and $s_{k}^{i}$ represents the $i$-th state in the $k$-th stage. $\alpha$, $\beta$ and $\gamma$ represent messages associated with the past, the present and the future, which can be calculated by forward and backward recursion. $\alpha(s_{k-1}^{i'})$ represents the forward probability of state $s_{k-1}^{i'}$. $\beta(s_{k-1}^{i'})$ represents the backward probability of state $s_{k-1}^{i'}$. $\gamma(s_{k})$ represents the message from received signal $Y$. $f_k$ is a virtual node, which validates the state transition of the trellis. If $f_k=1$, it represents a valid transmission, and $f_k=0$ otherwise. $\delta(\Bar{u}_k)$ records the probabilities of different SUM bit cases. Overall, BCJR decoding algorithm consists of four steps: initialization, forward and backward recursion, and SUM bit decision. We will now explain them in details.

\textbf{Initialization:} We consider reduced state joint trellis as a cycle-free Tanner graph. Our sum-product algorithm starts at the first and last stages. Since we use the zero-tail convolutional code, the initial and terminal states of different edge device’s convolutional encoders are zero state. Therefore, the forward probabilities $\alpha$ and backward probabilities $\beta$ in different states can be initialized as:

\begin{equation}
     \alpha (s_0^i) = 
     \begin{cases}
     1, & \text{if $i = 0$,}\\
     0, & \text{otherwise}.
     \end{cases}
\end{equation}

\noindent and

\begin{equation}
     \beta (s_k^i) = 
     \begin{cases}
     1, & \text{if $i = 0$,}\\
     0, & \text{otherwise}.
     \end{cases}
\end{equation}
where $i$ is the state index.

\textbf{Forward and Backward Recursion:} We first perform forward recursion from the $0$-th stage to the $K$-th stage to calculate $\alpha$ and $\gamma$, and then perform backward recursion from the $K$-th stage to the $0$-th stage to calculate $\beta$ and $\delta$.

We first introduce the forward recursion.
In the $k$-th stage, the branch metric $\gamma(\Bar{c}_k)$ is the difference between the received signal and the expected signal given the current state. Assuming BPSK modulation with a code rate of 1/2, we obtain the encoded bits $\Bar{C}_k=(C_{k,1}^A,C_{k,2}^A,C_{k,1}^B,C_{k,2}^B)$, where these channel-encoded bits correspond to the BPSK-modulated symbols $X^A[k]=(X_{2k-1}^A,X_{2k}^A)$ and  $X^B[k]=(X_{2k-1}^B,X_{2k}^B)$. Therefore, the branch metrics can be represented as
\begin{equation}
    \begin{split}
        \gamma (\Bar{c}_k) &\propto \prod_{n=2k-1}^{2k} \\
        &exp\left(\frac{-\Vert Y[n]-H^A[n]X^A[n]-H^B[n]X^B[n] \Vert^2}{2\sigma^2}\right),
    \end{split}
\end{equation}
which is actually the summation of the two likelihood probabilities of the encoded-bit pairs generated by source bits $S_k^A$ and $S_k^B$, and $2\sigma^2$ is the variance of noise.

After calculating the branch metrics, we can compute the forward probabilities $\alpha$ between linked states. As for the $i$-th state in the $k$-th stage, the forward probability $\alpha(s_k^i)$ can be represented as
\begin{equation}
    \alpha(s_k^i)=\sum_{\mathbf{L}\{i\}} f_k(s_{k-1}^{i'},S_k^F,\Bar{c}_k,s_k^i)\alpha(s_{k-1}^{i'})\gamma(\Bar{c}_k),
\end{equation}
where $\mathbf{L}\{i\}$ is a set of states in the $(k-1)$-th stage linked with the $i$-th state in the $k$-th stage.

Then, we perform backward recursion from $K$-th stage to $0$-th stage. The backward probability $\beta(s_k^i)$ of the $i$-th state in the $k$-th stage can be represented as:
\begin{equation} 
   \beta(s_{k-1}^{i'})=\sum_{i\in\mathbf{L}\{i'\}} f_k(s_{k-1}^{i'},S_k^F,\Bar{c}_k,s_k^i)\beta(s_k)\gamma(\Bar{c}_k),
\end{equation}
where $\mathbf{L}\{i'\}$ is a set of states in the $k$-th stage linked with the $i'$-th state in the $(k-1)$-th stage.

To decode the SUM bits, decoder requires the APP of different SUM bit result $S_k^F$ in every stage, which is determined by the APP of every state. We add the calculation of APP in backward recursion to reduce the decoding complexity. Specifically, for every linked state pairs $s_{k-1}^{i'}$ and $s_k^i$, we calculate the $\delta(s_{k-1}^{i'},s_k^i)$ after $\beta$ calculation. The APP $\delta(s_{k-1}^{i'},s_k^i)$ can be represented as
\begin{equation} 
   \delta(s_{k-1}^{i'},s_k^i)= f_k(s_{k-1}^{i'},\Bar{u}_k,\Bar{c}_k,s_k)\alpha(s_{k-1}^{i'})\gamma(\Bar{c}_k)\beta(s_k^i).
\end{equation}

\textbf{SUM Bit Decision:}
 Notice that we need APP of a stage instead of one state to determine the SUM bit. Therefore, we sum up $\delta(s_{k-1}^{i'},s_k^i)$ of all possible linked state pairs and obtain the probabilities of SUM bit results. %
 The SUM bit decision can be represented as 

\begin{equation} 
   S_k^F=\operatorname*{arg\,max}\sum_{s_{k-1}^{i'},s_k^i} \delta(s_{k-1}^{i'},s_k^i).
\end{equation}
We choose the SUM bit with the largest probability as the final decoding result.

\begin{figure}[t!]
	\centering
	\setlength{\abovecaptionskip}{0.cm}
	\includegraphics [scale=0.4,trim=0 0 0 0]{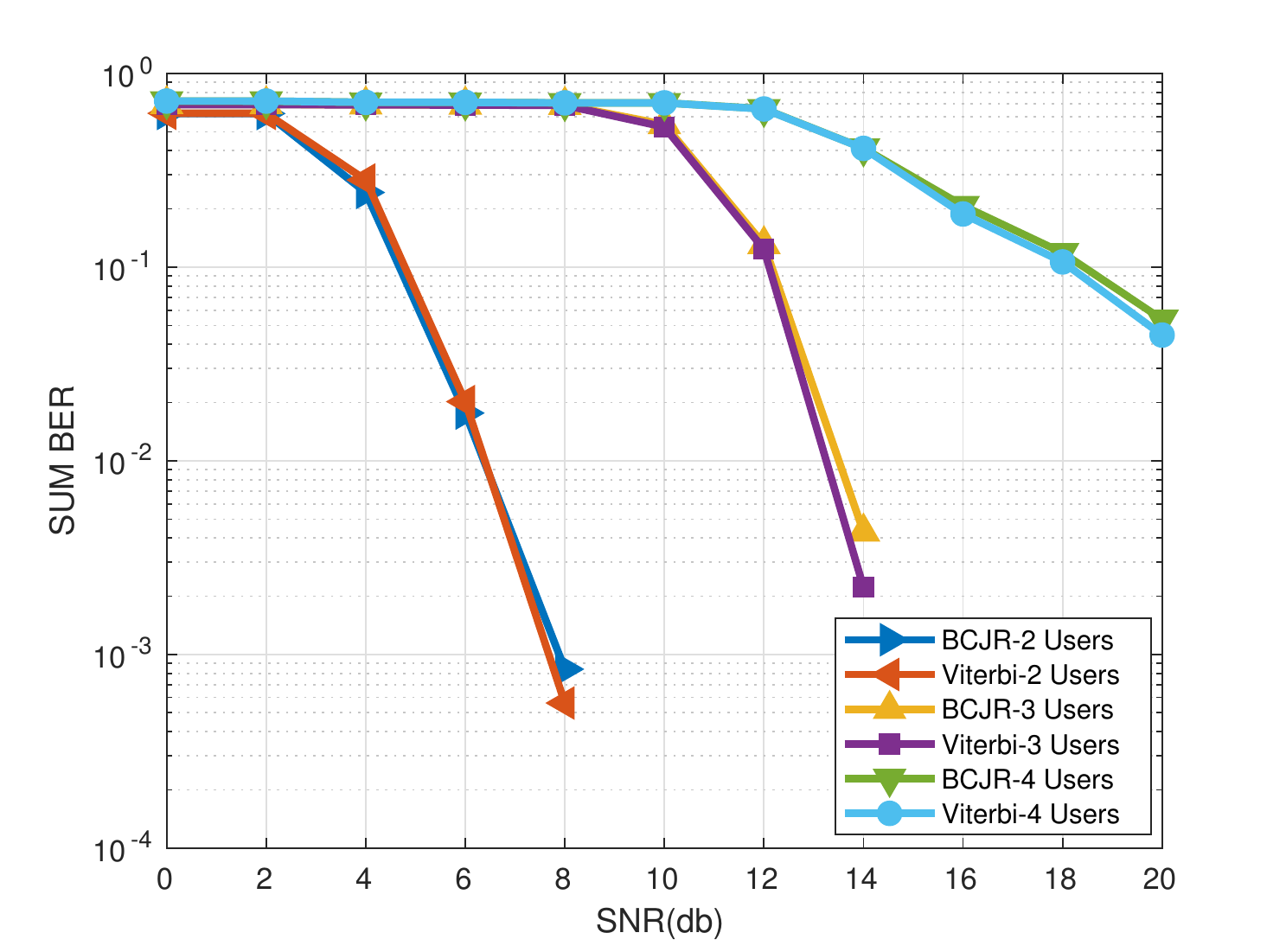}
	\caption{SUM BER versus SNR results for the BCJR and Viterbi algorithms under realistic channel on USRP.}
	\label{fig:usrp_bcjr}
\end{figure}

\subsection{Results}
We first perform simulations to investigate the decoding performance of the designed BCJR decoder.
The simulation conditions are identical to those specified in Section 6.1. We conduct separate tests for relative channel phase offest of $0$, $\pi/2$ and $\pi/4$ radian. FSJD retains all path metrics and states in each iteration, while RSJD selectively retains the 256 states with the minimum path metrics in each iteration. The forward recursion in the BCJR algorithm is performed synchronously with the path metric calculation in the Viterbi algorithm. BCJR-RSJD utilizes the 256 states recorded by Viterbi-RSJD in each iteration for the forward recursion calculation. After completing all iterations, BCJR-RSJD utilizes the recorded states to perform backward recursion and calculate the posterior probabilities. Fig.~\ref{fig:bcjr_viterbi} shows the the decoding performance of the BCJR and Viterbi algorithms in the case of two users. In the range of 4-10 dB, the simulation results show that BCJR is slightly better than Viterbi.

Then we perform experiments to investigate the decoding performance of the designed BCJR decoder under realistic channels. Fig.~\ref{fig:usrp_bcjr} shows the SUM BER results of BCJR and Viterbi for different decoders under a realistic channel measured on USRP. The data is identical to that specified in Section 6.1. Similar to the simulation results, BCJR is slightly better than Viterbi using the experiment data.

The above results turn out to be negative for the SUM bit-optimal decoder for convolutional codes. The proposed 
BCJR algorithm operates in the same trellis as the Viterbi algorithms (i.e., the approximated SUM codeword-optimal decoders), but it has a higher computation complexity due to message passing. However, their performances are similar. The phenomenon may be explained similar to the single-user communication system: the maximum a posteriori probability (MAP) decoder without a-priori information does not provide any performance gain against the maximum likelihood decoder (i.e., the Viterbi algorithm). The MAP decoder shows superior performance in the iterative decoder such as Turbo codes. We will study the decoding algorithm for Turbo-coded AirComp in the future.

}

\section{Parallel Single-User Decoders (PSUD)} \label{appendix:psud}
Parallel single-user decoder (PSUD) is another type of channel decoding and aggregation approach that leverages the multi-user decoding (MUD) technique. Different from joint decoders, PSUDs decode each device's data separately. Take two-user simultaneous transmissions for example. We first decouple the superimposed signal in each subcarrier to each device's symbol decoding likelihood
\begin{equation}
	Pr(Y[n]|C^A[n]) = \sum_{C^B[n]} Pr(Y[n]|C^A[n],C^B[n]),
\end{equation}
and
\begin{equation}
	Pr(Y[n]|C^B[n]) = \sum_{C^A[n]} Pr(Y[n]|C^A[n],C^B[n]).
\end{equation}
The decoupled symbol likelihoods are fed into two conventional single-user decoders to find the codeword-optimal codewords for convolutional codes or the bit-optimal codewords for LDPC codes. Then codewords are mapped to source bits for each user, and SUM bits can be computed.

\ifCLASSOPTIONcompsoc
\else
\fi

\ifCLASSOPTIONcaptionsoff
  \newpage
\fi

\bibliographystyle{IEEEtran}
\bibliography{IEEEabrv,bibfile}

\end{document}